%% file: main.tex
\documentclass[fleqn,usenatbib]{mnras}

\usepackage[T1]{fontenc}

\DeclareRobustCommand{\VAN}[3]{#2}
\let\VANthebibliography\thebibliography
\def\thebibliography{\DeclareRobustCommand{\VAN}[3]{##3}\VANthebibliography}

\usepackage{graphicx}	
\usepackage{amsmath}	
\usepackage{longtable}
\usepackage{hyperref}
\usepackage{color}
\usepackage{comment}
\usepackage{rotating}
\usepackage{caption}
\usepackage{ltablex}
\usepackage{tabularx,xltabular,capt-of}
\usepackage{txfonts}

\title[GRB progenitors revisited]{Gamma-ray burst progenitors revisited}

\author[A.~J. Levan et al.]{Andrew~J. Levan,$^{1,2}$\thanks{E-mail: a.levan@astro.ru.nl}
Jillian C. Rastinejad,$^{3}$\thanks{NASA Einstein Fellow}
Helena-Margaret S. Grabham,$^{4}$
Daniele B. Malesani,$^{1,5,6}$
\newauthor
Nial R. Tanvir,$^{7}$
Eric Burns,$^{8}$
Benjamin P. Gompertz,$^{9,10}$
Gavin P. Lamb,$^{4}$
Ashley A. Chrimes,$^{11,1}$
\newauthor
Peter G. Jonker$^{1}$,
Om Salafia,$^{12,13}$
Nikhil Sarin,$^{14, 15}$
Ilya Mandel,$^{16,17}$
Antonio Martin-Carrillo,$^{18}$
\\
$^{1}$Department of Astrophysics/IMAPP, Radboud University, 6525 AJ Nijmegen, The Netherlands\\
$^{2}$Department of Physics, University of Warwick, Coventry, CV4 7AL, UK\\
$^{3}$Department of Astronomy, University of Maryland, College Park, MD 20742, USA\\
$^{4}$Astrophysics Research Institute, Liverpool John Moores University, IC2 Liverpool Science Park, 146 Brownlow Hill, Liverpool, L3 5RF, UK\\
$^{5}$Niels Bohr Institute, University of Copenhagen, Jagtvej 128, 2200
Copenhagen, Denmark\\
$^{6}$Cosmic Dawn Center (DAWN), Denmark\\
$^{7}$School of Physics and Astronomy, University of Leicester, University Road, Leicester, LE1 7RH, UK\\
$^{8}$Department of Physics \& Astronomy, Louisiana State University, Baton Rouge, LA 70803, USA \\
$^{9}$School of Physics and Astronomy, University of Birmingham, Edgbaston, Birmingham, B15 2TT, UK\\
$^{10}$Institute for Gravitational Wave Astronomy, University of Birmingham,
Edgbaston, Birmingham, B15 2TT, UK\\
$^{11}$European Space Agency (ESA), European Space Research and Technology Centre (ESTEC), Keplerlaan 1, 2201 AZ Noordwijk, the Netherlands\\
$^{12}$INAF - Osservatorio Astronomico di Brera, Via E. Bianchi, 46, I-23807 Merate, Italy\\
$^{13}$INFN - Sezione di Milano-Bicocca, Piazza della Scienza 3, I-20126 Milano, Italy\\
$^{14}$Kavli Institute for Cosmology, University of Cambridge, Madingley Road, CB3 0HA, UK\\
$^{15}$Institute of Astronomy, University of Cambridge, Madingley Road, CB3 0HA, UK\\
$^{16}$School of Physics and Astronomy, Monash University, Clayton, Victoria 3800, Australia \\
$^{17}$OzGrav, Australian Research Council Centre of Excellence for Gravitational Wave Discovery, Australia \\
$^{18}$School of Physics and Centre for Space Research, University College Dublin, Belfield, Dublin 4, Ireland\\
}

\date{Accepted XXX. Received YYY; in original form ZZZ}

\pubyear{\the\year{}}

\newcommand{\totgold}{29~}
\newcommand{\goldlongduration}{21~}
\newcommand{\goldlmerger}{9~}
\newcommand{\goldlweak}{3~}
\newcommand{\goldcollapsar}{9~}
\newcommand{\goldshortduration}{8~}
\newcommand{\goldsmerger}{8~}
\newcommand{\goldsweak}{0}

\newcommand{\totsilver}{31~}

\newcommand{\silverlmerger}{4}
\newcommand{\silverlweak}{8}
\newcommand{\silvercollapsar}{6}

\newcommand{\silversmerger}{5}
\newcommand{\silversweak}{8}

\newcommand{\totbronze}{20~}
\newcommand{\bronzelmerger}{2~}
\newcommand{\bronzelweak}{2~}
\newcommand{\bronzecollapsar}{14~}
\newcommand{\bronzesmerger}{2~}
\newcommand{\bronzesweak}{0~}

\newcommand{\totallcats}{\textbf{80~}}

\begin{document}
\label{firstpage}
\pagerange{\pageref{firstpage}--\pageref{lastpage}}
\maketitle

\begin{abstract}
Recently, several long-duration gamma-ray bursts (GRBs) associated with kilonovae have cast doubt on the traditional, dichotomous mapping between $\gamma$-ray duration and progenitor system. Here, we investigate the rates and properties of bursts which appear to cross this dichotomy with a sample study of GRBs for which progenitor constraints are possible. We first build a sample of known {\em Swift}-detected GRBs at $z<0.3$, finding \goldshortduration short- and \goldlongduration long-duration GRBs. Of these long GRBs, we find \goldlmerger bursts with deep limits on supernova emission, evidence for kilonova emission, or association with a quiescent galaxy (31\% $\pm$ 9\% of all GRBs at $z<0.3$), implying that a significant fraction of nearby long GRBs likely do not come from massive stars. At $z<0.3$, no short GRB has an observed supernova counterpart. We find comparable numbers when expanding to $z<0.5$ and other $\gamma$-ray telescopes, though we obtain a decreased fraction of bursts with robust constraints on a progenitor.
We further find that the long GRBs with no associated supernovae possess on-average fainter afterglows and lie in less star-forming host galaxies than those with supernovae, supporting that these events may originate in compact object mergers. We estimate approximate volumetric rates, finding similar (on-axis) rates for short GRBs and supernova-less long GRBs of $\sim 0.5-2.5$ Gpc$^{-3}$ yr$^{-1}$, although a search for possible low-redshift hosts of the complete {\em Swift} catalog suggests that our sample may be $\sim 50$\% complete. If supernova-less long GRBs arise from compact object mergers, this implies that $\sim 30-70$\% of all $z<0.3$ {\em Swift} long GRBs may arise from mergers and that the $z<0.3$ rates of mergers from long and short GRBs are comparable. These findings hold substantial implications for gravitational-wave coincidence and heavy element enrichment. 
\end{abstract}



\section{Introduction}
First established over 40 years ago, the dichotomy in the duration distribution of $\gamma$-ray bursts (GRBs) \citep{mazets82,kouveliotou93} has been repeatedly confirmed by multiple $\gamma$-ray satellites. In each iteration, GRB durations, commonly defined as the period over which 90\% of the total fluence is received or $T_{90}$, break down into two classes: so-called ``short'' and ``long'' GRBs. The dividing time between the classes is typically drawn from the populations observed by the {\it Compton Gamma-Ray Observatory} ({\it CGRO}) Burst and Transient Source Experiment (BATSE) at  $T_{90} = 2$ seconds \citep{kouveliotou93}.   

It has long been assumed that this dichotomy maps directly into two distinct progenitor channels, with rather little overlap. The discovery of the unusual, broad-lined, stripped envelope (Type Ic-BL) supernova, SN~1998bw, in the error box of the equally unusual, and extremely under-luminous GRB 980425 \citep{galama98} identified a potential massive star progenitor connected to long GRBs.
This was eventually cemented by the spectroscopic detection of the supernova SN~2003dh, that exhibited remarkably similar properties to SN~1998bw in the afterglow of GRB~030329 \citep{hjorth03,stanek03}. To date, tens of long GRB and Ic/Ic-BL supernova pairs have subsequently been identified \citep[for a review see, e.g.][]{Cano+17_review}. These supernovae are generally interpreted under the collapsar model \citep{woosley93}, in which a rapidly-rotating massive star undergoes core-collapse. This creates both a centrifugally supported disc that powers a beamed, ultra-relativistic jet and GRB as well as a comparatively isotropic supernova. The observed properties are also plausibly explained by black hole accretion powered stellar collapse \citep[see e.g.][for a comprehensive list]{fryer99}, or even magnetar driven explosions \citep{Metzger11,mazzali14, metzger15}. 

For short GRBs, the path to a secure progenitor was longer. The first suggestions of an origin in neutron star (NS) mergers pre-date afterglow discoveries by more than a decade \citep[e.g.][]{blinkonov84,pacczynski86,eichler88}. Their presence in old populations was apparent from the discovery of the first afterglows \citep{gehrels05,bloom05,barthelmy05}. Larger samples have revealed diversity in host stellar population properties \citep{fong13,oconnor22,fong22,nugent22}, and that short GRBs are often offset large distances from their putative hosts \citep{fong13b}, sometimes to a degree that host identification was non-trivial \citep{berger13,tunnicliffe14}. Such properties were entirely consistent with the expectations of GRBs born from the merger of two compact objects \citep{eichler88,pacynzski98}, in which the combination of mass loss and natal kicks to neutron stars provides spatial velocities of tens to hundreds of km s$^{-1}$ to the binary. As gravitational wave (GW)-induced merger times can span from $10^7$ to $>$10$^{10}$ years, this provides a natural explanation for high offsets from the centers of galaxies \citep{kalogera04,church11,eldridge19,mandhai22,gaspari24b,gaspari25, Mandel_2026}. 

Further progress towards identifying the progenitors of short GRBs  was made by searching for the expected 
explosive signatures of  merger events. In particular, it was long-theorized that compact object mergers containing neutron stars would yield neutron-rich outflows capable of rapid neutron capture (the ``$r$-process''; \citealt{lattimer77,rosswog98,rosswog03,thielemann17}). These captures form isotopes far from the valley of stability, whose subsequent radioactive decay should power a faint, rapidly-fading transient \citep[e.g.][]{pacynzski98,kulkarni05,metzger10,metzger12}. The general picture somewhat parallels supernovae in long GRBs, except that the signatures expected in short GRBs are roughly ten times faster and ten times fainter. These $r$-process-powered transients are referred to as kilonovae or macronovae in the literature\footnote{In this paper we use the kilonova terminology from here on.} and are a distinctive signature of a compact object merger. Early searches for kilonovae in short GRBs were unsuccessful \citep[e.g.][]{hjorth05a, bloom05}. However, these searches concentrated predominantly on the optical light. Significant progress was propelled by studies showing that $r-$process elements, in particular the lanthanides, produce significantly higher opacities than iron group elements, yielding redder transients compared to supernovae \citep{tanaka13,barnes13}. Searches following this realisation had immediate success in identifying a likely kilonova in GRB 130603B \citep{tanvir13,berger13}. Subsequent searches suggested that similar signatures may be present in more short GRBs \citep[e.g.][]{jin16,gompertz18,troja18,lamb19,troja19,ascenzi19,rossi20,rastinejad21}, although such inferences are challenging given the faintness of short GRBs and the difficulty of separating their afterglow and kilonova light \citep{wallace_sarin}.  

Ultimately, that at least some short GRBs arise from compact object mergers was sealed with the detection of both a short GRB (GRB 170817A) and a GW chirp (GW170817) from a merging neutron star binary \citep{abbott_bns,abbott_multimessenger,goldstein17,savchenko17}. This event also provided the best studied kilonova to date, AT2017gfo \citep[e.g.][]{arcavi17,Chornock+17,kasliwal17,smartt17,soares-santos17,tanvir17}. The total of several hundred photometric observations \citep{villar18} outnumbers the sum of those in all other events combined, while the rich spectral series continues to provide new insight into the details of kilonovae \citep[e.g.][]{watson19,hotokezaka23,sneppen23,sneppen24,gillanders26}.

These observations make clear that {\em some} short GRBs arise from mergers and {\em some} long GRBs arise from collapsing stars, producing supernovae. Moderate overlap in the two populations, such that bursts near the divide between long and short bursts may arise from either population \citep[see e.g.][]{bromberg13}, has always been apparent. However, once one moved substantially from that boundary, it was expected that bursts should be dominated by one progenitor population. For example, under the assumption that the intrinsic distribution for both long and short GRBs is log-normal, we might expect {\em Fermi}-GRBs to have a 10\% contribution from the short population at $T_{90}=10$s, but likely $<1\%$ by 100s \citep[e.g. Figure 4 in][]{vonkienlin20}. 

Recent observations have strongly challenged this scenario. Firstly, in the short GRB population, the identification of a supernova in GRB 200826A, with a duration of 
$T_{90} = 1.136 \pm 0.132$, implied that some collapsars can have duration substantially shorter than the notional two second divide \citep{ahumada21,rossi22}.
Secondly, two recent GRBs with durations $>30$ s (GRBs 211211A and GRB 230307A) have been robustly associated with a kilonova. These detections are both photometric \citep{rastinejad22,yang22,troja22,yang23,levan24a}, via the detection of strong-IR excesses and in one case spectroscopic via late-time JWST observations \citep{levan24a,gillanders25}. These observations suggest that previously seen long GRBs without associated supernovae \citep{fynbo06,dellavalle06,gal-yam06,levan23b} can be readily interpreted as mergers, and support the 
tentative indications of kilonova signatures that have been suggested in some of these events \citep{jin15,yang15,jin21,Rastinejad+25,stratta25}. 

The logical question that arises from these observations is the relative purity of samples of short and long duration GRBs. \textit{To what degree do these two populations genuinely represent distinct progenitor channels?} At a basic level this question casts doubt on the textbook model for the mapping of progenitors to $\gamma$-ray duration. More broadly, the answer to this question poses several other potential ramifications. For example, long GRBs powered by compact object mergers challenge the standard picture in which the jet duration is related to the mass of accreting material, requiring new theories as to the relationship between the jet and ejecta mass or the central engine , or alternative merger progenitors \citep[e.g.][]{gottlieb23,king07}. Further, estimates of the rates of compact object mergers or of heavy element enrichment often make assumptions about the short GRB rate (e.g., \citealt{Rouco+23,Fishbach+26}). A substantial additional contribution
from long GRBs could require significant revisions in enrichment calculations, especially if some of these events arise from short delay time systems which have previously been ascribed as supernovae.  Alternatively, long GRBs are frequently invoked as tracers of massive stars and the star formation rate (SFR) , potentially into the era of reionisation. In using long GRBs to study distant galaxies \citep[e.g.][]{tanvir12,schady24} or the fraction of photons escaping from massive stars to reionise the Universe \citep{tanvir19}, the assumption is that their lines of sight point directly to massive stars. If this is not the case, the insight they provide may be incorrect \citep[e.g.][]{levan25a}.

Furthermore, the ``merger" population does not necessarily just contain the mergers of two neutron stars as seen in GW170817 \citep{abbott_bns}. Black hole neutron star mergers have also long been considered as important possible contributors to the GRB population \citep[e.g.][]{rosswog05,davies05,troja08,gompertz20}. More recently, the concept of white dwarf - black hole or white dwarf -- neutron star mergers has also come to the fore. Although these were first suggested to explain supernova-less GRBs more than twenty years ago \citep{king07}, they have also been invoked as a possible origin for both GRB 211211A \citep{yang22,zhong23} and GRB 230307A \citep{yang23,chen24}, and may naturally have similar galactic locations \citep{chrimes25}. Although such mergers were not readily expected to synthesise heavy elements, increasing work has been investigating them as possible GRB progenitors \citep[e.g.][]{liu25} that would also contribute to the compact object merger population. 

Despite these substantial implications, the challenges of distinguishing between different progenitors have limited the progress towards understanding their overlap. It is appealing to utilise multiple observational diagnostics to better hone the nature of the progenitor of any given burst. Indeed, a comprehensive framework for such classification was presented by \cite{zhang09}, who, following the classification of supernovae suggested a split into Type I (non-collapsar) and Type II (collapsar) GRBs, based on several different observational signatures. Alternately, classification schemes using machine learning trained on the limited simple of bursts with known progenitors have been attempted (e.g., \citealt{DimpleArun23,Negro+25, 2026Maccary, 2026Zhang}), although could benefit substantially from additional bursts with clean progenitors. Such approaches clearly have great value, and represent the best classifications that can be undertaken with limited information. However, they can be challenging to interpret in the frequent cases where full information is not available, and can risk mis-classification of perhaps the most interesting cases (e.g. collapsars at large offsets \citep{blanchard16}, or mergers with very short delay times, e.g.,  \citealt{beniamini24A}). Hence, such approaches are perhaps most valuable when applied to samples of objects, rather than individual events. 

Here, we address these open questions by considering the sample of bursts observed by the {\em Neil Gehrels Swift Observatory} ({\em Swift}; \citealt{Gehrels04}) at low redshift ($z<0.3$ and $z<0.5$). At these distances, supernovae can be robustly excluded by mid- to large-aperture ground-based telescopes. From this sample we attempt to determine the relative fractions of the observed population which (1) may originate from massive stars, (2) where a massive star origin is ruled out and a compact object merger is most likely, and (3) those for which the origin remains uncertain. 

In Section~\ref{sec:methods} we outline our different samples and make constraints on their progenitors. In Section~\ref{sec:core_props} we use our sample to examine other discriminating properties between progenitors such as their prompt emission, afterglow, host galaxy properties. In Section~\ref{sec:rates} we comment on potential selection effects in our sample and provide simple rate estimates for long GRBs. We further discuss the implications of this sample in Section~\ref{sec:discussion}. Finally, we present our conclusions in Section~\ref{sec:conclusion}. Throughout, we report all magnitudes in the AB system and assume a standard $\Lambda$CDM cosmology  \citep{planck18}.

\section{Methods}
\label{sec:methods}

We define and justify a set of criteria to identify a uniformly-selected sample of 
low redshift
GRBs for which we may investigate their
origins in stellar collapse or merging compact objects.
We elect to use the detection or deep limits on a GRB supernova counterpart, an unambiguous indication of stellar collapse, as our primary discriminant between the two progenitors. This is because supernovae are longer-lived and a factor of $\sim 100$ more luminous compared to kilonovae. Thus, the majority of GRBs at $z < 0.5$ have some constraints on the presence or otherwise of an accompanying supernova. We discuss additional properties we use to discriminate the progenitors in Section~\ref{sec:obs_constraints}.

We acknowledge the possibility that supernova-less GRBs may not originate in compact object mergers. A massive star association for supernova-less GRBs does remain plausible in some cases if, for example, there is heavy, but unrecognised extinction, or if the collapse of the massive star ejected very little mass. However, in these cases, and given the identification of other GRBs with kilonova and compact object mergers, it is more economical (i.e.\ does not require an additional progenitor) to consider a merger origin. We further discuss implications of this assumption and alternate possibilities in our forthcoming companion paper (Rastinejad et al. in prep.).

\begin{figure*}
	\centering
		\includegraphics[width=0.85\textwidth]{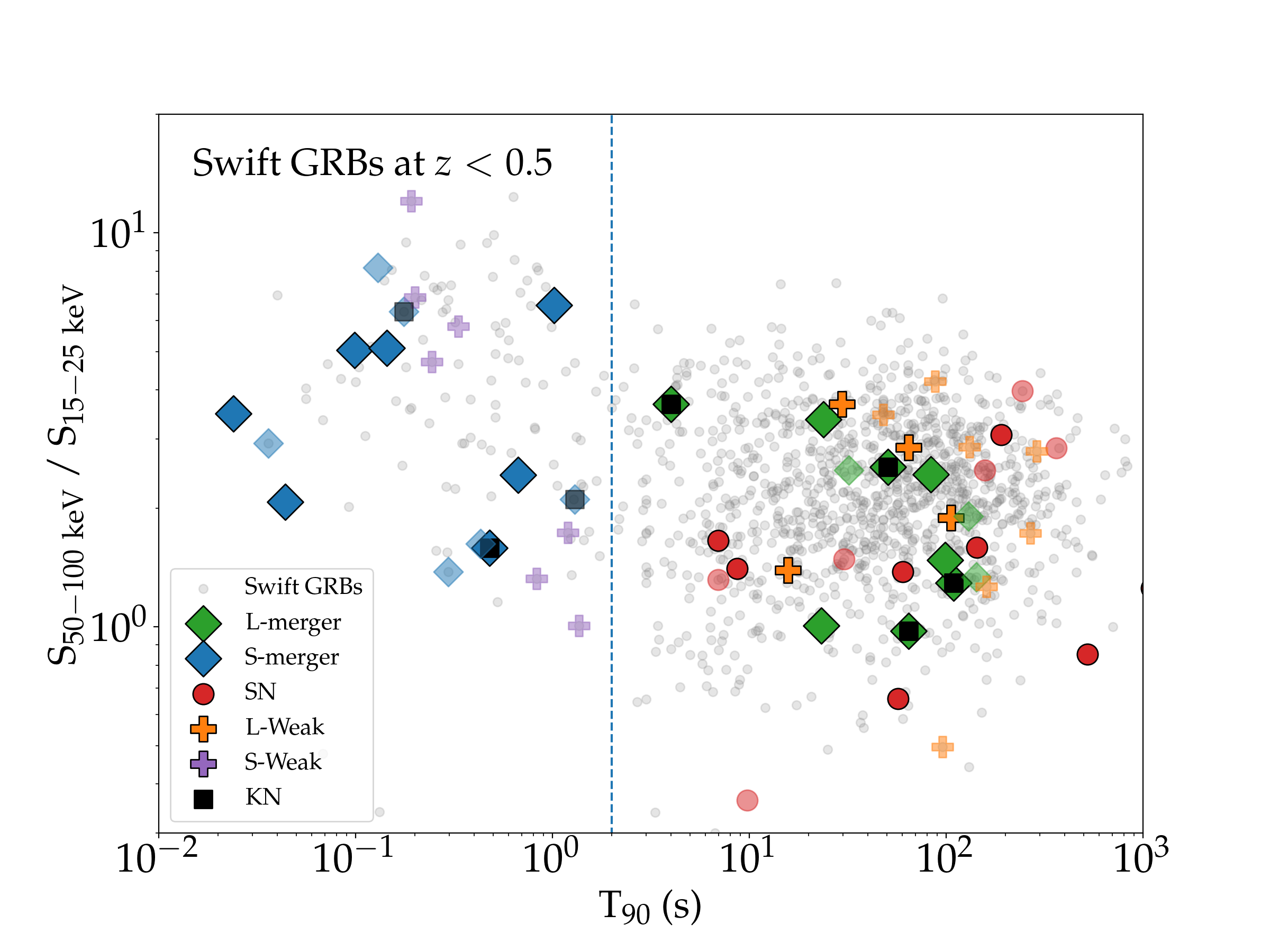}
	\caption{Hardness duration diagram for {\em Swift} GRBs in our sample at $z<0.5$. The background points indicate the whole 
    {\em Swift} population. The location of bursts at $z<0.3$ are indicated in bold, while those at $0.3 < z < 0.5$ are more transparent versions of the same sample symbol. The proposed progenitor types for each burst are marked.}
	\label{hardness}
\end{figure*}

\subsection{The GRB sample}
\label{sec:sample}

To robustly identify or place deep limits on supernova counterparts to GRBs and to determine their rates, we require cuts based on redshift and $\gamma$-ray detection mission. Large-aperture telescopes may identify GRB supernovae spectroscopically at $z\lesssim 0.3$ and photometrically at $z\lesssim0.5$, motivating two cuts based on supernova detection robustness. The use of large-aperture telescopes for follow-up has most frequently been applied to GRBs with X-ray afterglows detected by \textit{Swift}'s X-ray Telescope (XRT; \citealt{burrows05}) as these provide localizations of just $\sim$several arcseconds. Furthermore, the {\em Swift} sample is large, and detected by a single $\gamma$-ray detector. We therefore consider the \textit{Swift}/XRT sample of GRBs to be our primary source of bursts for consideration. 

We define three samples of GRBs: one of \textit{Swift}/XRT-detected bursts at $z<0.3$, a second more inclusive sample of \textit{Swift}/XRT-detected bursts at $z<0.5$, and a final selection of GRBs detected by any mission and known to be at $z<0.5$. We do not include Fast X-ray Transients (FXTs) in our analysis, including SN\,2008D \citep{soderberg+08,Mazzali2008,Modjaz+09} and those found by the recently-launched Einstein Probe \citep{Yuan+22}, unless these FXTs were also detected in $\gamma$-rays. We discuss implications for this choice in Section~\ref{sec:discuss_othertransients}.

To determine what bursts fall into each category requires knowledge of each burst's redshift. There is no definitive list of GRB redshifts, but we search both GCN circulars and the refereed literature to provide as complete a list as possible. 
We note that obtaining redshifts in absorption for low redshift GRBs is often challenging because the strongest lines lie in the rest-frame UV and are not in the optical window. Hence, distance measurements are necessarily based mostly on the host galaxy associations.
For simplicity here, and for consistency with the past work of \citet{fong22}, we accept a redshift where the probability of chance alignment is $P_{cc} < 0.1$ \citep{bloom02}. We discuss implications of this method in Section~\ref{subsec:sample_sel_bias}. 

Our criteria result in \totgold $z<0.3$ GRBs and an additional \totsilver $z<0.5$ GRBs detected by {\em Swift} from 1 Jan 2005 to 31 December 2024 (20 years; Table~\ref{tab:numbers_by_cat}). We do not place further constraints on the GRB properties (for example on their peak flux, or fluence) given the limited sample size. Extending beyond {\em Swift} detection, we find a further find \totbronze GRBs (\totallcats GRBs across all categories), including several foundational events from the pre-{\em Swift} era. We mostly focus our subsequent analysis on the {\em Swift} detected samples as the differing selection functions for prompt emission and varied follow-up strategies render robust inference of the heterogeneous but all encompassing bronze sample more challenging. We list the samples in Table~\ref{sample}, and show the hardness duration diagram for the ({\em Swift}) population in Figure~\ref{hardness}. 

\subsection{Constraints on GRB Progenitors}
\label{sec:obs_constraints}

We employ several criteria
in constraining the progenitor of each GRB. Our starting point
is to separate the sample into three categories: (i) GRBs for which we identify a clear photometric and spectroscopic detection of a supernova counterpart, indicating a collapsar origin, or conversely, a clear  detection of a kilonovae, indicating a merger origin, (ii) GRBs for which such a determination cannot be made due to weak or no constraints, and (iii) GRBs for which deep imaging reveals no evidence of a supernova to limits substantially fainter than the proto-type GRB supernova, SN~1998bw (e.g., \citealt{galama98}), likely indicating a merger origin. 

\begin{figure*}
    \centering
    \includegraphics[width=16cm]{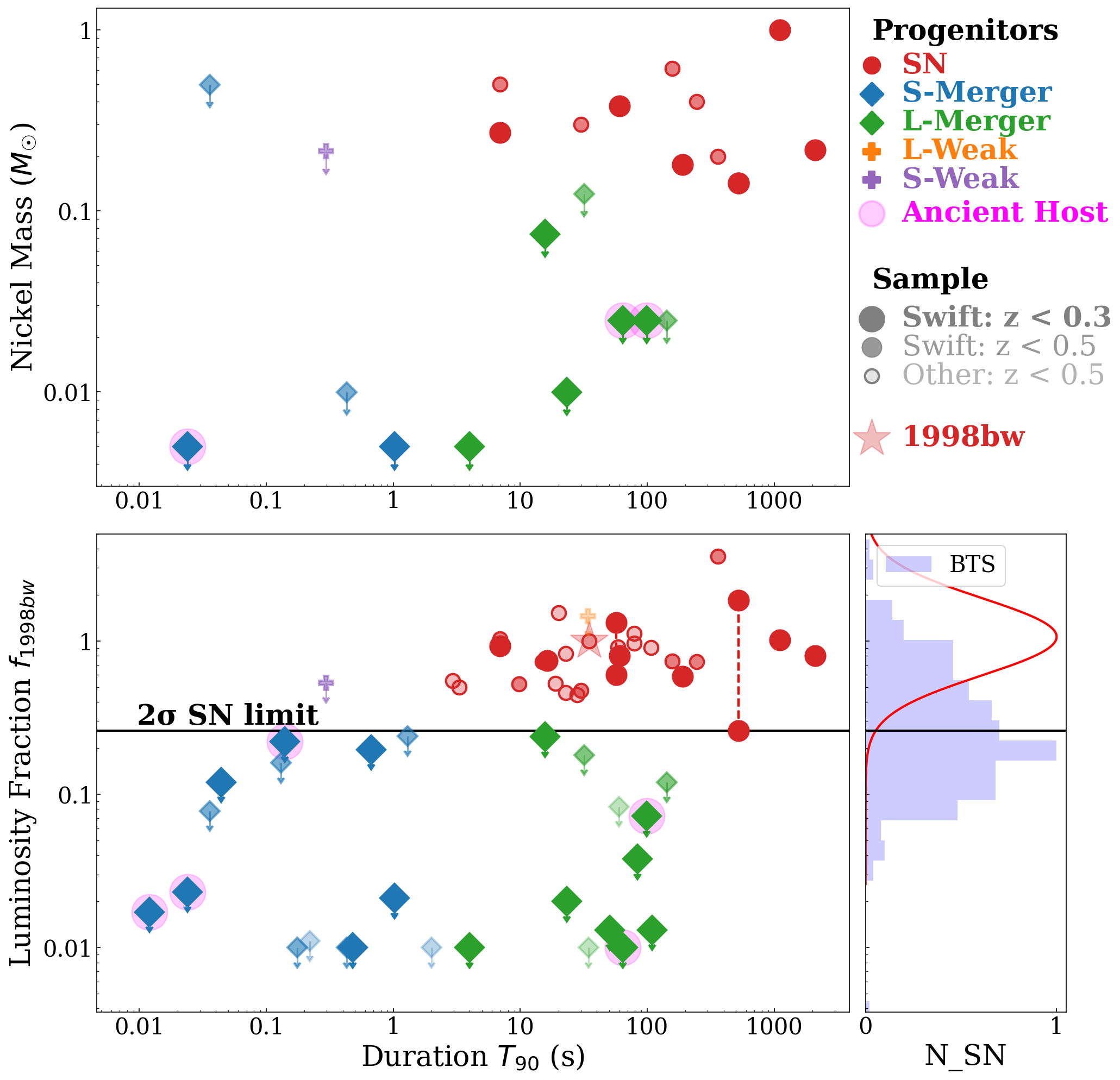}
    \caption{Supernova properties shown to vary over GRB duration time $\text{T}_{90}$(s). \textit{Top panel:} Nickel masses, $M_{Ni}$ ($M_\odot$) vs duration from the sample of GRBs at $z<0.3$ and $z<0.5$. 
    \textit{Bottom Panel:} Supernova brightness measured as a fraction of SN 1998bw vs GRB duration. Vertical red dashed lines indicate the range of values corresponding to GRBs 100316D and 190829A. All scales are represented logarithmically. Points corresponding to the {\em Swift} samples are displayed respectively with upper limits given by downward facing arrows. The legend shows  markers associated with the five burst groups highlighted above. The horizontal black line corresponds to the luminosity limit for a supernova according to this sample and is constrained using a gaussian fit to be 5\% at  $f_{1998bw}$ = 0.26. GRB 980425 associated with SN 1998bw is represented with a star marker. The histogram attached to luminosity scatter represents the probability density function of Type Ic supernovae from the Bright Transient Survey. There is a clear discontinuity between BTS data and the sample.}
    \label{fig:nickelandlum}
\end{figure*}


To this end, we utilize the luminosity ratio of a given GRB optical/IR counterpart detection or upper limit to SN 1998bw at the same rest-frame time and wavelength, and define this as $f_{\rm 1998bw}$. We gather the available photometry of SN\,1998bw \citep{Vincenzi_2019} and construct a synthetic, representative light curve using the \verb|sn1998bw_template_with_extrapolation| function within the {\tt Redback} package \citep{Sarin_2024}. We then interpolate over this SN\,1998bw light curve and produce a luminosity at a redshift-corrected wavelength and epoch corresponding to each actual GRB observation (Table~\ref{sample}). For validity we repeat this interpolation process using {\tt MOSFiT} \citep{mosfit} and find the two codes produce comparable results. We note that such approaches are approximate, and we do not seek to extract the properties of the supernovae in any detail by fitting full multi-wavelength light curves, since for many well studied events this has been undertaken in detail, and our determination of supernove presence is generally supported based on these claims in the literature. Rather, our aim is to use the same approach to obtain an approximate luminosity for the supernova, but most critically to place limits where one is not seen. Since most well-studied GRB supernovae show only a modest dispersion in peak magnitude and light curve morphology (e.g., \citealt{Cano+17_review}), there is a reasonable justification to cast our limits in the light of SN~1998bw. To determine whether we can confidently rule out a supernova counterpart for GRBs, we fit a log-normal distribution to the $f_{\rm 1998bw}$ values for GRBs with confirmed and potential supernovae (Figure~\ref{fig:nickelandlum}). We use the 2-$\sigma$ lower limit of $f_{\rm 1998bw}$ = 0.26 from the fitted log-normal distribution as our criteria to distinguish between GRBs with robust and weak constraints on a supernova counterpart.

We show the values of $f_{\rm 1998bw}$ in the bottom panel of Figure~\ref{fig:nickelandlum}, from photometric evidence outlined in Table \ref{sample_photometry}. Strikingly, all of the bursts for which supernovae are seen have relatively luminous supernovae, comparable (within a factor of $\sim 2$) to SN~1998bw. We also plot the probability density function of Type Ic supernovae from the ZTF Bright Transient Survey (BTS; \citealt{perley20}), highlighting that the majority of upper limits are not only much deeper than the expected brightness of SN~1998bw, but rule out a significant fraction of all core-collapse supernovae. This disconnect with the observed ZTF BTS data displayed in the histogram emphasises that the population of long GRBs with deep limits on supernova emission are unlikely to be the product of massive stars. 

To further assess the robustness of our constraints we compile estimates of $^{56}$Ni mass from \cite{2016MNRAS.458.2973P, 2019MNRAS.487.5824A, Finneran_2025} and employ the most constraining limit for each event to measure an upper bound on a $^{56}$Ni mass (see Appendix~\ref{subsec:ni56_limits} for method). We apply this to the GRBs in the $z<0.3$ and $z<0.5$ samples and show the results in the top panel of Figure~\ref{fig:nickelandlum}. We observe a bifurcation between measurements and upper limits on $^{56}$Ni mass, further supporting these are two distinct progenitor populations.  

Beyond constraining the presence of a supernova, we consider the presence of a kilonova counterpart and ongoing star formation within the host galaxy in determining the progenitor. A burst that arises from an quiescent galaxy should not be created by a massive star. There are several apparently ancient galaxies in the short GRB population, and at least two (GRBs\,050219A, 191019A) in the long GRB population. Further, there is a small but important set of bursts for which a kilonova has been claimed.  This comprises GRBs\,060614 \citep{jin15,yang15}, 130603B \citep{tanvir13,berger13}, 150101B \citep{troja19}, 160821B \citep{lamb19,troja19}, 191019A \citep{stratta25}, 211211A \citep{rastinejad22,troja22} and 230307A \citep{levan24a,yang23}. However, some caution should also be exercised in this sample since the level of confidence of the kilonova association is highly variable. Further, we caution that much of our data is compiled from individual burst analyses. Different works may undertake various methodologies with regard to, for example, supernova and host galaxy fitting. Hence, while the broad properties of the bursts can be compared (e.g. supernova or none, star-forming or quiescent host galaxy) caution should be used in comparing details.

Using this information we define five different categories of bursts to consider:
\begin{enumerate}
    \item ``Collapsars'': Long GRBs with supernovae (red in figures).  
    \item ``L-merger'': Long GRBs without supernovae ($f_{\rm 1998bw} < 0.26$) and/or in quiescent/ancient galaxies (green in figures).  
    \item ``L-Weak'': Long GRBs with weak supernova constraints ($f_{\rm 1998bw} > 0.26$; orange in figures). 
    \item ``S-Weak'': Short GRBs with weak supernova constraints ($f_{\rm 1998bw} > 0.26$; purple in figures).  
    \item ``S-merger'': Short GRBs without supernovae and/or in ancient/quiescent hosts (blue in figures). 
\end{enumerate}

\begin{figure}
	\centering
        \includegraphics[width=\columnwidth,angle=0]{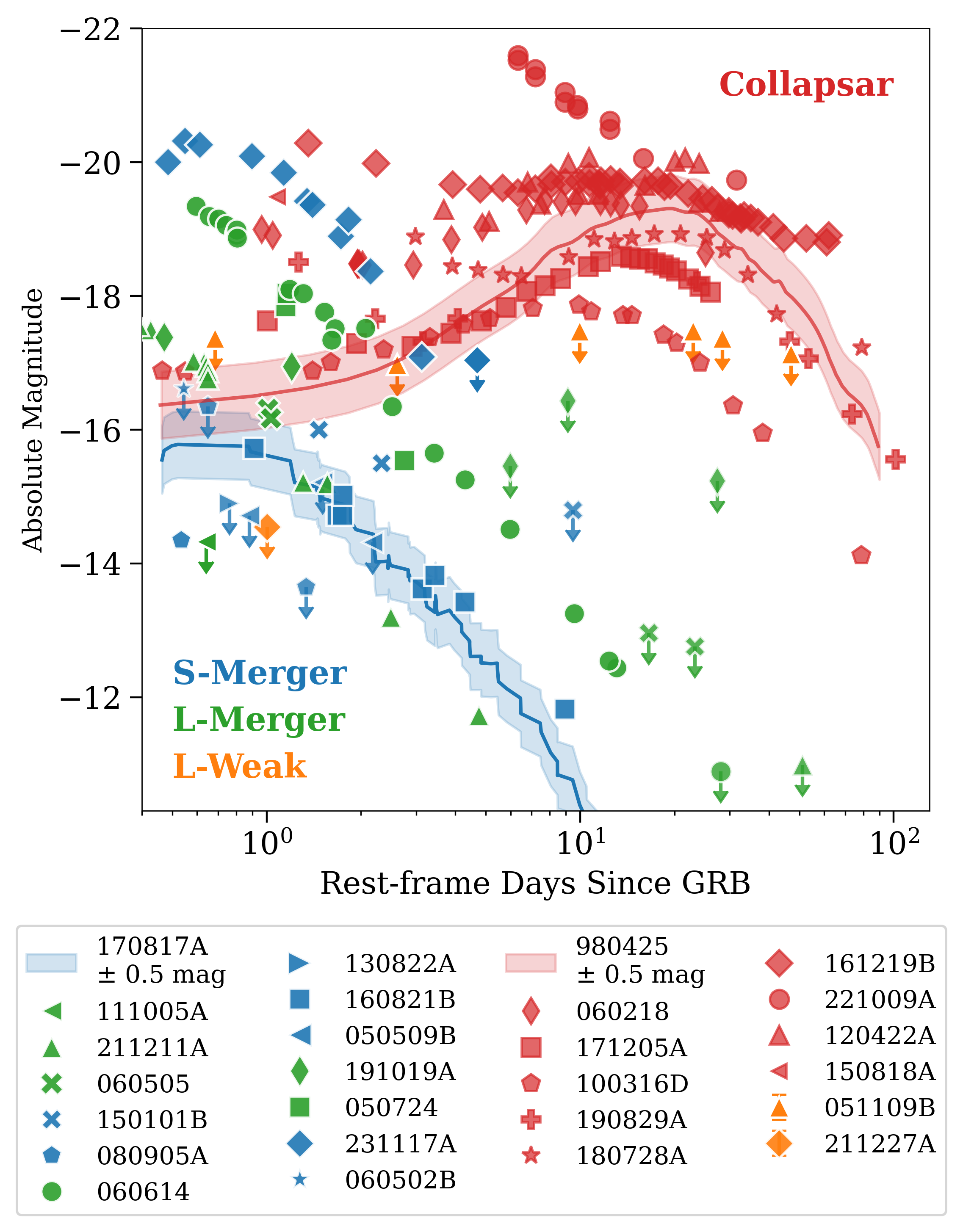}
	\caption{Optical detections and upper limits on the supernova and kilonova counterparts to our $z<0.3$ sample of GRBs (\textit{Swift} GRBs at $z<0.3$; \citealt{bloom05,hjorth05,berger05,Malesani+07,Kocevski+07,fynbo06,Chornock+10,rowlinson10,Melandri+12,jin15,Cano+17,michalowski18,troja18,izzo19,lamb19,troja19,hu21,rastinejad21,rastinejad22,Laskar+23,levan24a,rastinejad24a,Schroeder+25,Rossi+26}). We plot light curves in the $\approx$rest-frame $g/r$-band where available. We also plot AT\,2017gfo, the kilonova of GRB\,170817A/GW170817, and SN\,1998bw, the supernova of GRB\,980425 \citep{galama98,Clocchiatti+11,arcavi17,coulter17,Cowperthwaite+17,Drout+17,Diaz+17,Hu+17,Kasliwal+17,lipunov17,Pozanenko+17,Shappee+17,smartt17,soares-santos17,tanvir17,Utsumi+17,Valenti+17,villar18}.}
	\label{fig:kne-sne_1panel}
\end{figure}

\begin{table}
    \centering
    \caption{Summary of the numbers of GRBs by sample and progenitor category (Section~\ref{sec:methods}).}
    \begin{tabular}{c|ccc}
    \hline \hline
        Category & {\em Swift}, $z\leq0.3$ & {\em Swift}, $0.3 <z\leq0.5$ & non-{\em Swift}, $z<0.5$  \\
         & (\# GRBs) & (\# GRBs) & (\# GRBs)  \\
        \hline \hline
        Collapsar & \goldcollapsar & \silvercollapsar &  \bronzecollapsar \\
        L-Merger & \goldlmerger & \silverlmerger & \bronzelmerger  \\
        L-Weak & \goldlweak & \silverlweak &  \bronzelweak \\
        S-Merger & \goldsmerger & \silversmerger &  \bronzesmerger \\
        S-Weak & \goldsweak & \silversweak &  \bronzesweak \\
        \hline
        Total & \totgold & \totsilver & \totbronze \\
        \hline \hline
    \end{tabular}
    \label{tab:numbers_by_cat}
\end{table}

We report the numbers of bursts in each category per sample in Table~\ref{tab:numbers_by_cat} and show compiled rest-frame $\approx R$-band light curves, where available, for the $z<0.3$ sample in Figure~\ref{fig:kne-sne_1panel}. At $z<0.3$, we find \goldcollapsar collapsar bursts with spectroscopic evidence for a supernova (though see \citealt{levan23a,blanchard24} for GRB 221009A). We find a significant sample of long GRBs with no evidence for a supernova (\goldlweak L-Weak and \goldlmerger L-Merger bursts). The L-Merger bursts at $z<0.3$ are GRBs 050219A, 050724A, 051109B, 060505,  060614, 111005A, 191019A, 211211A and 211227A. For GRB 051109B supernova limits have not been reported previously, but we recover a moderately deep limit on a supernova ($F < 0.15 f_{\rm 1998bw}$) based on serendipitous observations of the field with the Catalina Sky Survey (see Appendix). In the $z<0.5$ sample, we are unable to confidently ascertain the presence of a supernova for a larger fraction of bursts. This is partly because redshifts of these GRBs are often discovered on longer timescales, and the larger luminosity distance means that supernova exclusion requires observations with larger facilities (Table~\ref{sample}). While short GRBs with supernovae are known to exist (e.g., \citealt{ahumada21,rossi22}) this group is not populated  at $z<0.5$. 

\begin{figure}
	\centering
		\includegraphics[width=\columnwidth]{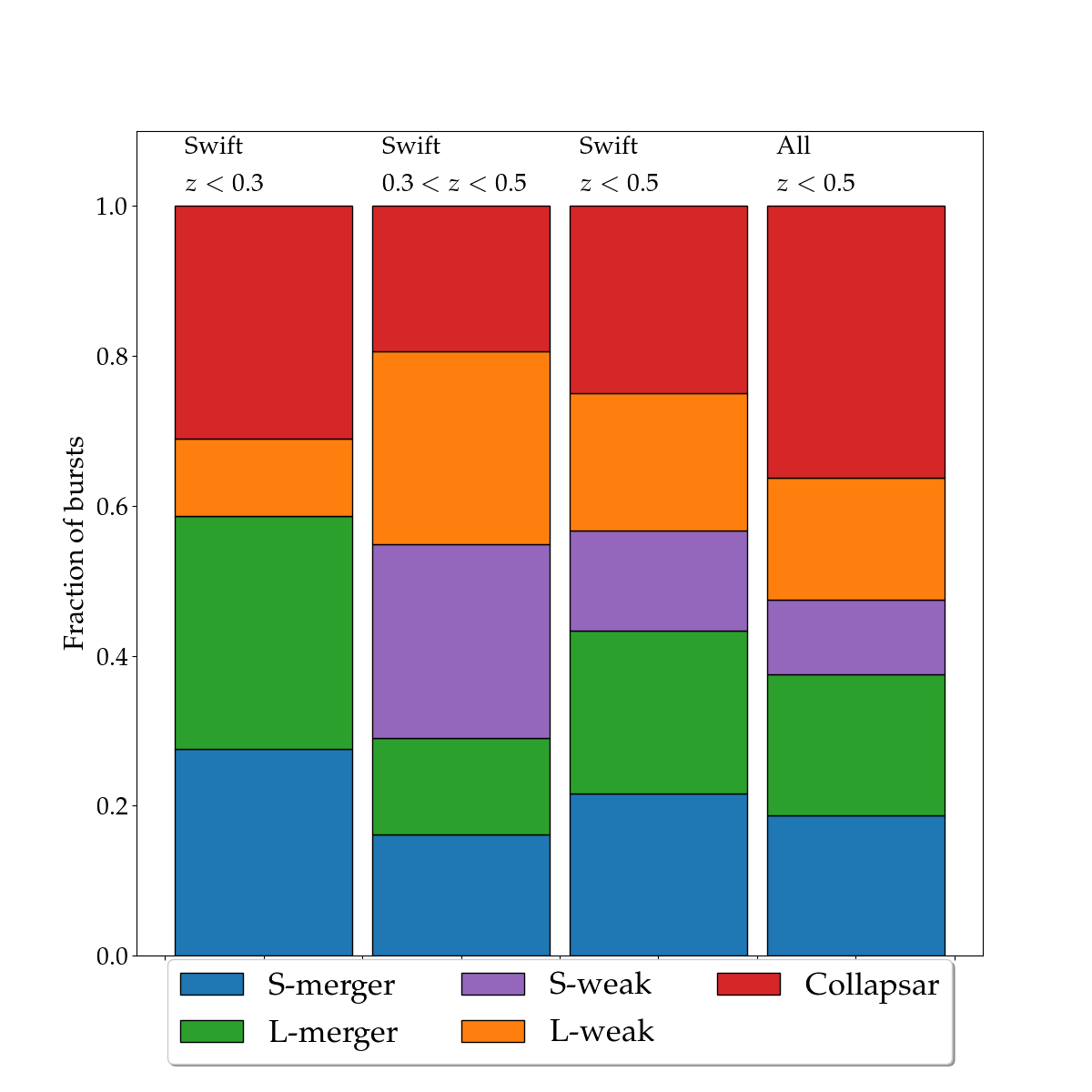}
	\caption{Fractions of GRBs in different classes based on different samples of comparison. GRBs are split according to the 5 different burst groups. Our primary sample ({\em Swift} $z<0.3$) has the largest fraction of events which lie in the likely long-merger category ($T_{90} > 2$s, no-supernova or ancient hosts). This may arise from a combination of better limits at low redshift, and genuine source rate evolution in higher$-z$ samples. In increasing our sample size the fraction of events with weak constraints increases. All samples are consistent with equal fractions of $\sim 20-60$\% of long GRBs arising from non-supernova channels. Note that the rates here have been set to equality at $z=0.15$, but this is intended to be illustrative of the trends given the uncertainty in both the observable type fractions and also the conversion from observed fraction to volumetric rate. }
	\label{type_frac}
\end{figure}

The resulting type fractions are shown in Figure~\ref{type_frac}, broken down into the different samples we consider. As can be seen, the observed fraction of GRBs with and without supernovae (including L/S-Weak bursts) is similar and consistent across the different selections, although we do note the larger fraction of collapsar bursts amongst the non-{\em Swift} population.

\begin{figure*}
	\centering
		\includegraphics[width=.49\textwidth]{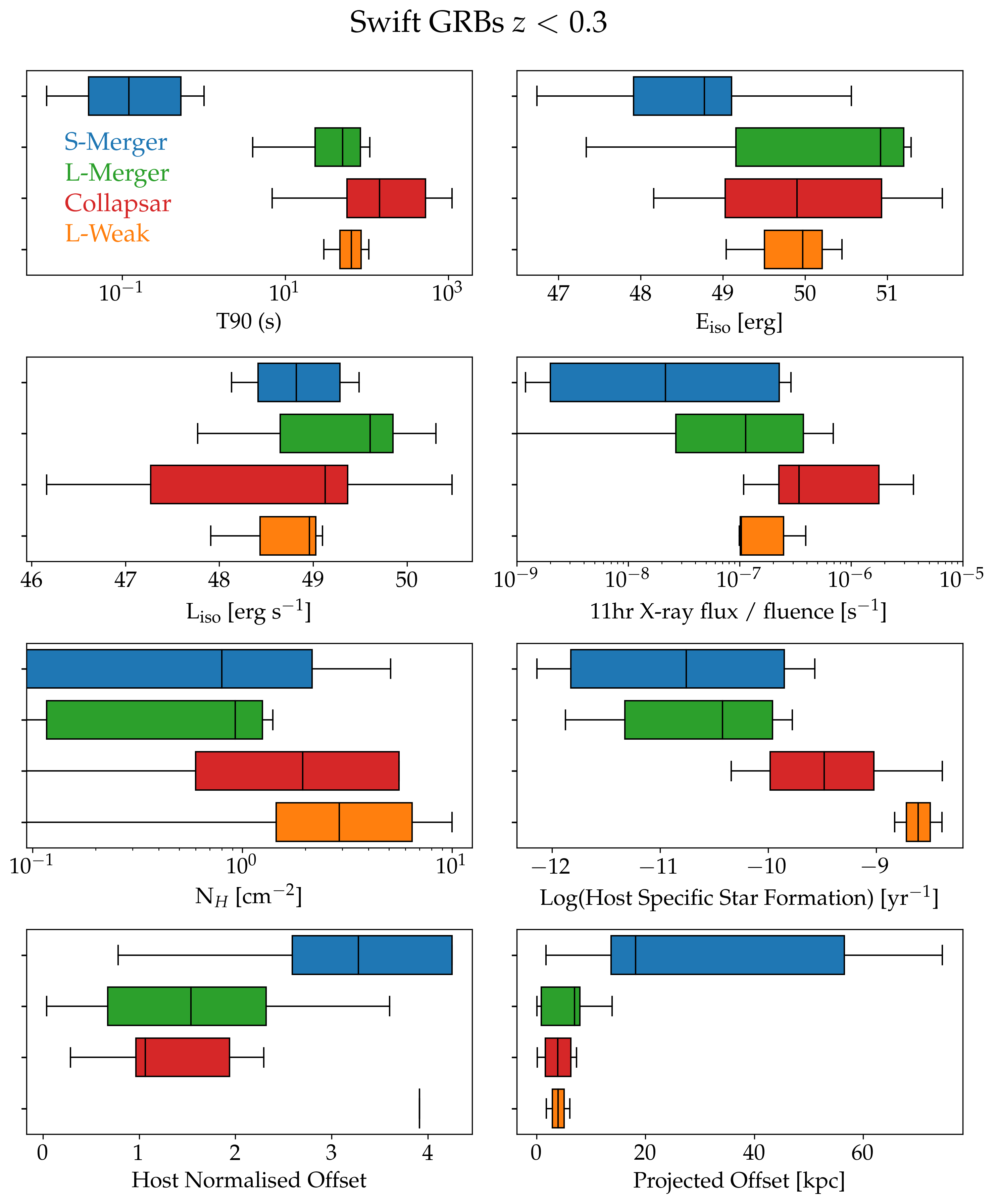}
        \includegraphics[width=.49\textwidth]{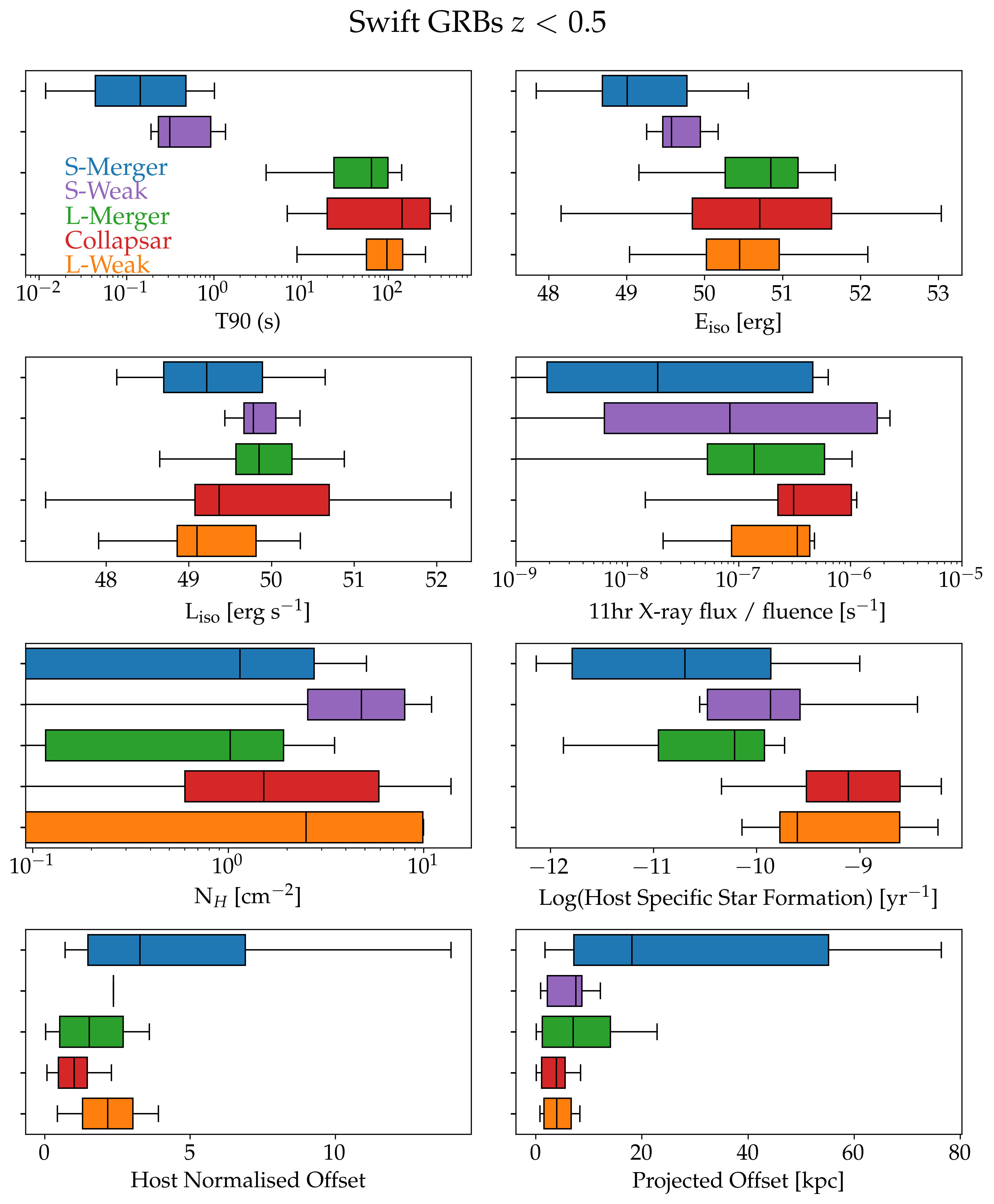}
	\caption{Box plots displaying the core parameters of interest in distinguishing between merger and collapsar GRBs for the $z<0.3$ sample (left) and $z<0.5$ sample (right). The vertical center line, colored box, and horizontal black lines denote the median, 50 percent and 99.7\% confidence intervals in each parameter, respectively. While the strongest distinctions are unsurprisingly based on duration, it is also apparent that the collapsar population extends to higher $L_{iso}$, especially at higher redshift, while the merger related events appear to have, on average somewhat fainter afterglows for their fluence, occur in galaxies of lower specific SFR and are offset at larger distances.  While the majority of these trends are not highly significant they may be useful in building a physically motivated picture for a given burst progenitor.}
	\label{gold_sample_box}
\end{figure*}

\section{Comparison of sub-sample properties}
\label{sec:core_props}

Beyond the strong distinction that arises via the presence (or absence) of a supernova, there are numerous observational properties that may differ between massive star and compact object merger progenitors. Here, we briefly consider several and show them graphically in the box-plots in Figure~\ref{type_frac} and Figure~\ref{gold_sample_box}. We pull the prompt emission properties of GRBs from the {\em Swift}-Burst Alert Telescope (BAT; \citealt{Swift_BAT}) catalog (\citealt{lien16}; and its online updates). 

\subsection{Duration and spectral hardness} 

We first compare the properties of the short and long GRBs at $z<0.3$ and $z<0.5$ to the overall \textit{Swift} population. Short GRBs make up 28\% and 35\% of each of our redshift-limited samples, respectively. This is a larger fraction of short GRBs than is present in the overall 2004-2024 {\em Swift} population, of which only 8.5\% are short. 
However, since the typical redshifts of long and short GRBs are very different, with short GRBs at a median redshift of $\bar{z} \sim 0.64$ \citep{nugent22} and long GRBs at $z > 2$ \citep{jakobsson06}, the outcome in the restricted horizon of $z<0.3$ is unsurprising. 

We show the location of the bursts in the hardness-duration plane compared to the bulk population in Figure~\ref{hardness}. Overall the bursts are similar to the bulk populations, with no indication that the we are preferentially sampling (or omitting) any region of this parameter space. We further discuss the implications of this substantial fraction of L-Merger/L-Weak bursts in our companion paper (Rastinejad et al. in prep.)

\subsection{Energetics \& Prompt Emission Relations}

We next investigate burst energetics and prompt energy relations. We first find the isotropic-equivalent energy ($E_{\rm iso}$) and istropic-equivalent luminosities ($L_{\rm iso}$). We show their median, 50 percent and 99.7\% intervals in Figure~\ref{gold_sample_box}. We find no strong differences in the energetics between the populations of long GRBs that show supernova signatures and those that do not. At low redshift, the collapsar GRBs are less energetic than L-Mergers, likely reflecting the presence of so-called low-luminosity GRBs amongst the collapsar population. Indeed, perhaps the most striking feature of the energetic distributions, in particular in terms of $L_{\rm iso}$, is the extremely broad range shown in the collapsar population which include (both by several orders of magnitude) the most and least energetic events.
Several population studies have tried to reconstruct the luminosity function of short GRBs, finding a parallel conclusion that there are fewer very luminous short GRBs compared to long GRBs \citep[e.g.][]{wanderman15,salafia23}.
Within our sample, the lowest $E_{\rm iso}$ and $L_{\rm iso}$ bursts tend to be at lower redshift, which is unsurprising given that they would drop below the detection threshold
at higher $z$. The shortest-duration events also typically have lower $E_{\rm iso}$ than the longer events, although they often have higher peak luminosities over a 1 s or 20 ms time-frame. 

We further study the relationship between a GRB's peak energy ($E_{\rm peak}$) and $E_{\rm iso}$, the so-called Amati relation \citep{amati06}, which is known to discriminate between collapsar and merger GRBs (or, as often presented, long and short GRBs). In the Amati relation, long GRBs fall along a fairly tight relation between $E_{\rm iso}$ and $E_{\rm peak}$, while short GRBs lie at high $E_{\rm peak}$ values above the relation. The Amati relation holds across multiple telescopes with spectral ranges capable of capturing $E_{\rm peak}$ (e.g., \textit{Fermi}, \textit{Konus}; \citealt{Tsvetkova2017ApJ,vonkienlin20}). Notably, however, {\em Swift}'s  spectral region (15-350 keV) is too narrow for $E_{\rm peak}$ constraints in the majority of cases, preventing us  exploring the Amati relation for all bursts in our samples.  

\begin{figure}
	\centering
		\includegraphics[width=\columnwidth,angle=0]{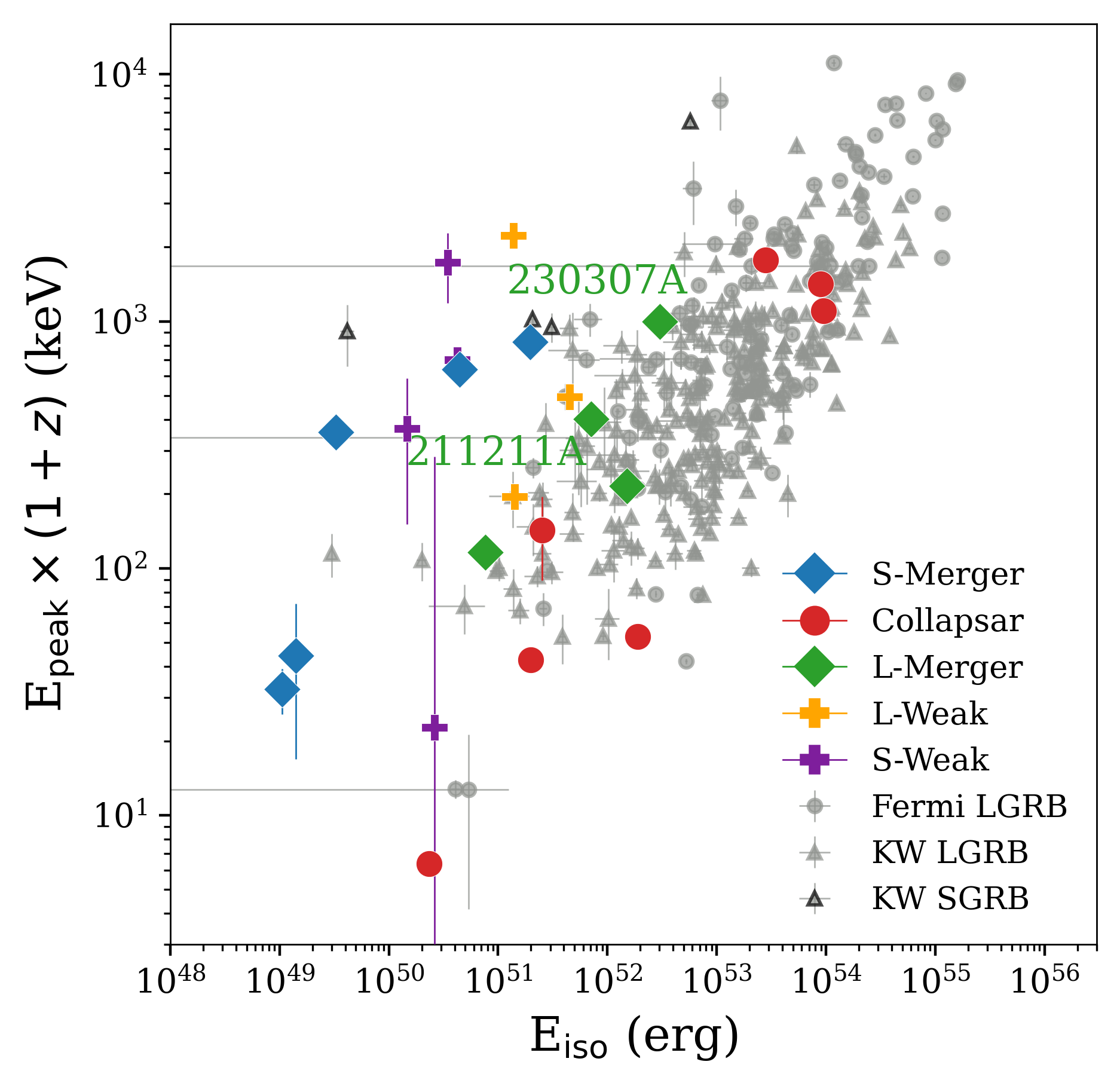}
	\caption{Isotropic-equivalent energy $E_{\rm iso}$  and peak energy $E_{\rm peak}$ for $z<0.5$ {\em Swift} GRBs also detected by Fermi (circles; \citealt{vonkienlin20}) or Konus-Wind (triangles; \citealt{Tsvetkova2017ApJ,Tsvetkova2021ApJ}). We also plot additional Fermi and Konus-Wind GRBs with known redshifts (grey) in the background and GRB 230307A. As expected, long GRBs from collapsars trace the Amati relation, while S-Merger and S-Weak bursts sit at lower $E_{\rm iso}$ and higher $E_{\rm peak}$ values. L-mergers and L-Weak bursts appear to straddle the divide between the collapsars and S-Mergers, adding further support to the L-Weak bursts originating in merger-like progenitors.}
	\label{fig:amati}
\end{figure}

In Figure~\ref{fig:amati}, we plot the redshift-corrected $E_{\rm peak}$-$E_{\rm iso}$ relation for {\em Swift} $z \leq 0.5$ GRBs also detected by \textit{Fermi} or \textit{Konus} against all GRBs with known redshifts seen by these telescopes. As expected, the collapsar GRBs lie along the Amati relation, while the S-Merger and S-Weak events reside at higher $E_{\rm peak}$ values compared to the relation. Interestingly, the L-Weak and L-Merger bursts, including GRB 211211A and GRB 230307A, lie at the edge of the collapsar Amati relation, in the direction of S-Merger and S-Weak GRBs. This adds further support to a compact object merger origin for at least some of the L-Merger and L-Weak bursts. Fundamentally, the typically low-energy nature of the bursts means they reside in sparsely population region of the Amati relation, making robust inference challenging. 

An additional prompt diagnostic that has been suggested to strongly discriminate between progenitors is the spectral lag \citep{gehrels06}. In particular, the merger-produced GRBs appear to have zero spectral lag in their prompt emission, whereas in the collapsar bursts the hard emission leads the softer emission. More concretely, the collapsar GRBs trace a so-called lag-luminosity relation, in which the spectral lag is longer for bursts with lower luminosities. Within our $z<0.3$ sample, we can only measure the spectral lag for nine bursts (GRBs 050724, 060502B, 060614, 130822, 161219B, 180728A, 191019A, 211211A, 211227A). Most of these are of the short GRB variety where there are sufficient features within the light curves to measure meaningful lags. For the longer bursts, which are often detected at a lower signal-to-noise ratio, the spectral lags are more difficult to measure. Subsequently, measured lags are only available for two bursts with supernovae (GRB 161219B, $\tau = -32 \pm 192$ ms and GRB 180728A, $\tau=4.0 \pm 42$ ms). Since both of these are consistent with zero, we conclude that within this sample there is a limited diagnostic value to the spectral lag, at least based only on the {\em Swift} observations.

\subsection{X-ray and Optical Afterglow Luminosities} 
\label{sec:ag_brightness}

\begin{figure}
	\centering
		\includegraphics[width=\columnwidth,angle=0]{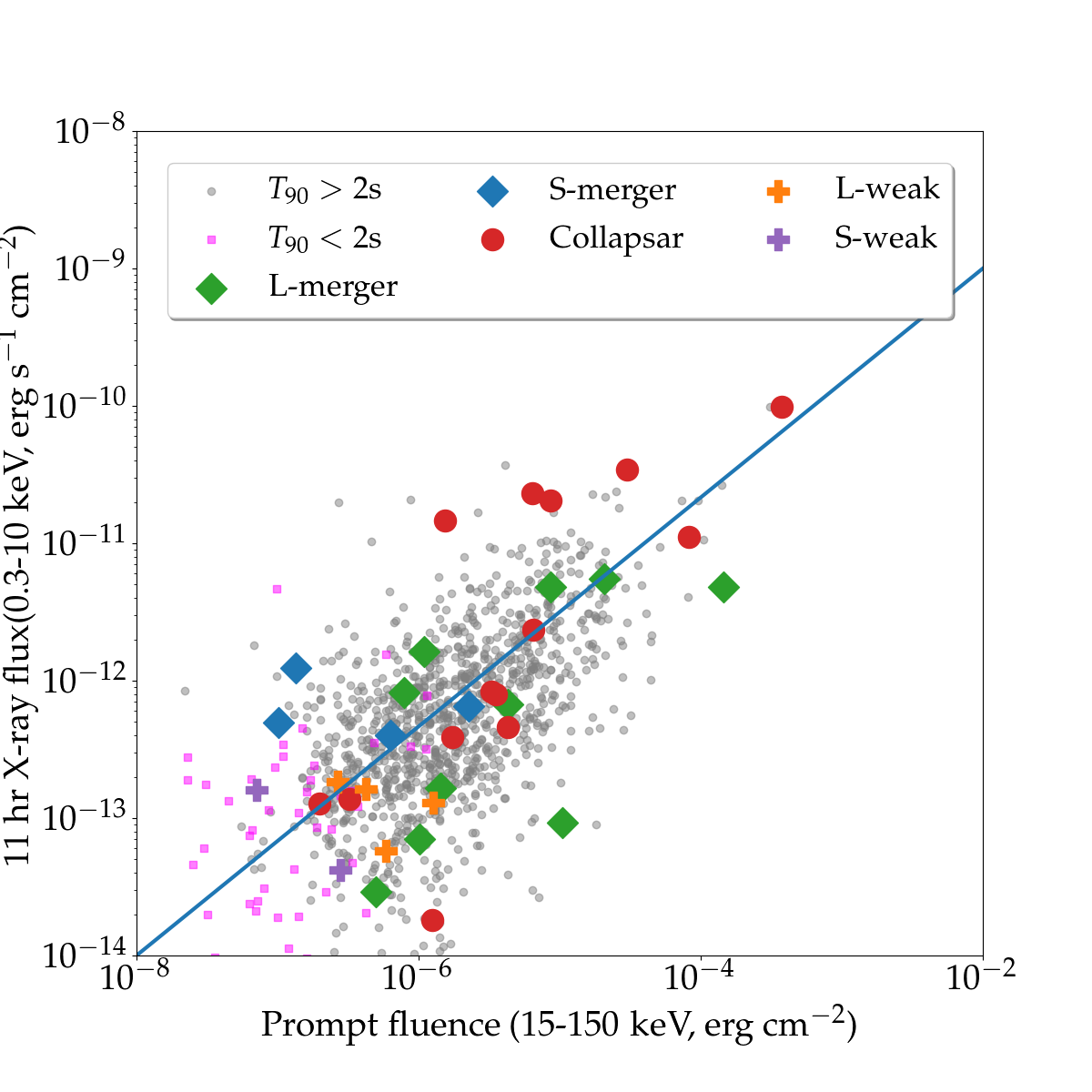}
	\caption{Correlation between prompt $\gamma$-ray fluence and X-ray afterglow brightness at 11 hours \citep{gehrels09,nysewander09}. Both long and short duration GRBs apparently show a broad correlation in which brighter prompt emission leads to a brighter afterglow. The best fit relation has a slope close to unity such that the two parameters are in direct proportion, although it is apparent that there is significant scatter of approximately 0.5 dex (a factor of 3). GRB 211211A and GRB 230307A are striking outliers in this relation, apparently lying well below the bulk of the population. This may be related to low circumburst densities given the galactic locations of the bursts, or could be intrinsic to the burst properties. There is a clear systematic offset that bursts without supernovae or in ancient host galaxies lie on, or below this relation, but seldom above it.}
	\label{xray_11hr_fluence}
\end{figure}

In general and with substantial scatter, {\em Swift} GRBs show a correlation between the prompt fluence and the X-ray and optical afterglow brightness 
\citep{gehrels08,nysewander09}. The overall slope of this relation is close to direct proportionality. It has previously been noted that
the long GRBs for which kilonovae are identified lie well below this relation, with both GRB 211211A and, in particular, GRB 230307A far off the sample of $>1000$ {\em Swift} GRBs
\citep{levan24a,yang23}. In Figure~\ref{xray_11hr_fluence} we show the X-ray flux of the gold sample at 11 hours post-burst, drawn from the {\em Swift} catalog\footnote{\url{http://swift.gsfc.nasa.gov}}. We note that in many cases (especially for the less fluent GRBs) this is the result of an extrapolation of the early afterglow decay, and not an interpolation on a well sampled light curve. For some bursts the final detections occurred during periods of rapid decay, and hence these fluxes are extremely low. It is unclear if these values are physically reasonable. 

Bursts with strong supernova limits, in ancient galaxies, or with associated kilonovae are preferentially offset below the 1:1 line such that they, on average, have fainter afterglows than expected for their prompt fluence (Figure~\ref{xray_11hr_fluence}). Several events (e.g. GRBs 191019A, 211211A, 230307A) are strong outliers in this relation, while others lie within the broader distribution of bursts. However, the faint X-ray afterglow appears to be a feature of supernova-less GRBs. Since broadly speaking X-ray afterglow brightness scales (weakly) with the density of the ambient medium this would be consistent with their interpretation as compact object mergers which have traveled away from their dense birth sites to regions within their galaxies where the ambient densities are lower. 

\begin{figure}
	\centering
        \includegraphics[width=\columnwidth,angle=0]{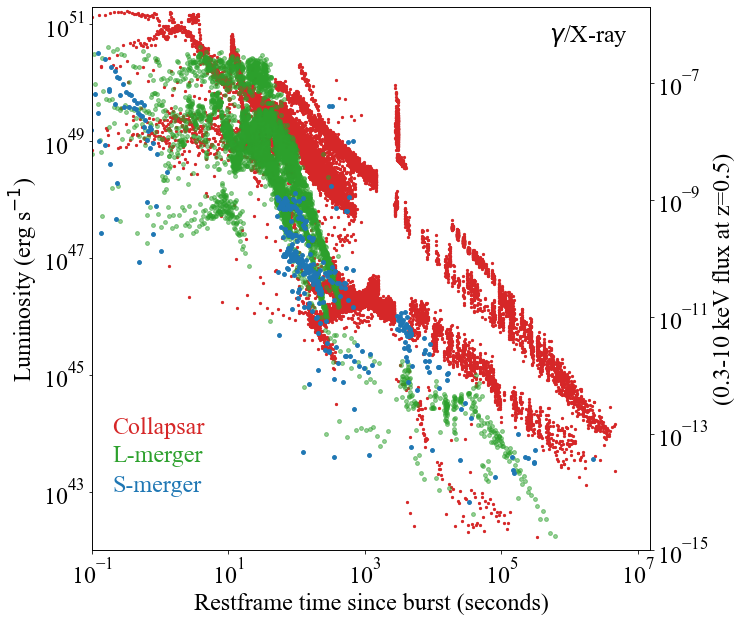}
        \vspace{-0.2cm}
        \includegraphics[width=\columnwidth,angle=0]{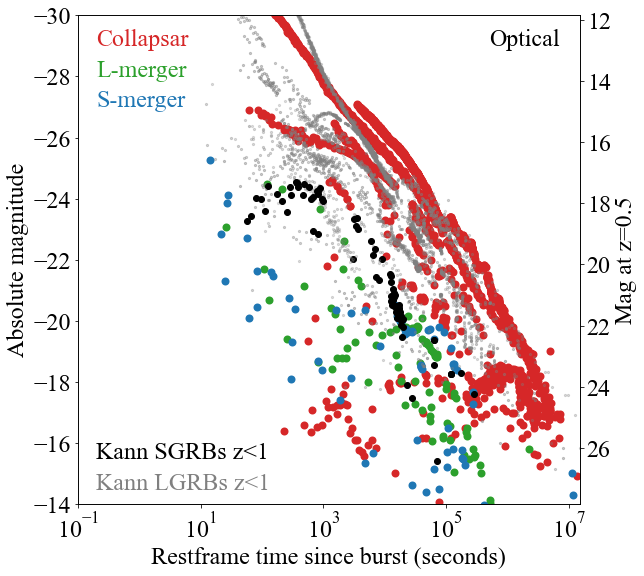}
	\caption{The X-ray (top) and optical (bottom) afterglows of GRBs witin our sample, plotted in luminosity and aboslute magnitude space. It is apparent that in both regimes the the collapsar population spans a very wide range of brightness, especially at late times. This likely reflects the low-luminosity and normal populations of GRBs, with higher luminosity events prevalent at higher redshift. In contrast the suggested merger populations are, on average, lower luminosity events without the extension to higher luminosity observed for the collapsars. This suggests that it will be more difficult to locate and study the afterglows of merger GRBs, and that, especially as one moves to higher redshift, a flux limited sample will be dominated by collapsar events. }
	\label{fig:xray_opt_aglcs}
\end{figure}

We can also extend this analysis into the optical regime. It has previously been noted that short GRBs appear to have fainter optical afterglows than long bursts \citep{kann+11}. In Figure~\ref{fig:xray_opt_aglcs} we show the Kann plot of our population (where optical observations are available from \cite{kann+11,kann24,Dainotti24}), along with a similar plot for the X-ray luminosities based on the {\em Swift} Burst Analyzer light curves from the UKSSDC \citep{evans07,evans09}. This confirms, that even in the same redshift regimes the merger afterglows are, on average, substantially fainter than the collapsar bursts. Figure~\ref{fig:xray_opt_aglcs} also highlights that the collapsar events diverge into two populations: the low-luminosity population of events such as GRB/XRF 060218, which are dominated in the optical by supernova-like emission at most epochs, and the more luminous events (such as GRB 221009A) which are afterglow dominated. The lack of bright afterglows for merger-driven GRBs appears to be a consistent feature across both the short and long-duration bursts. This could potentially be reflective of differences in the immediate environments of the bursts, or in the explosion mechanisms themselves. The difference may well have important implications for the detectability of the afterglows (rather than just the bursts) at higher redshifts, and we return to this later in Section~\ref{sec:rates}.

\subsection{X-ray absorption} 

In principle the X-ray absorption should also show indications of the environment in which the burst is formed, with higher columns expected along lines of sight to regions of star formation. Our sample of local bursts with a known supernova might result in selection bias as supernovae cannot be seen in regions with very high absorption which also manifests as optical extinction \citep{schady07,schady10}. To obtain hydrogen column densities we report PC mode measurements of the $N_H$ to GRBs taken at later times (since during the early afterglow phases spectral evolution can complicate $N_H$ measurements). We obtain these measurements from the {\em Swift} archive maintained at the UKSSDC \citep{evans07,evans09}, and for cases where the redshift is not known within the catalog, rebuild products at the appropriate redshift. The mean hydrogen column density ($N_{\rm H}$) for the supernova-less long GRBs is lower than that for the collapsar GRBs, and there is a notable absence of high-$N_{\rm H}$ events in the supernova-less sample (Figure~\ref{gold_sample_box}). However, given the large uncertainties, we are unable to conclude there is a significant difference in the subcategories.

\subsection{Host galaxy properties}

\begin{figure}
	\centering
		\includegraphics[width=\columnwidth,angle=0]{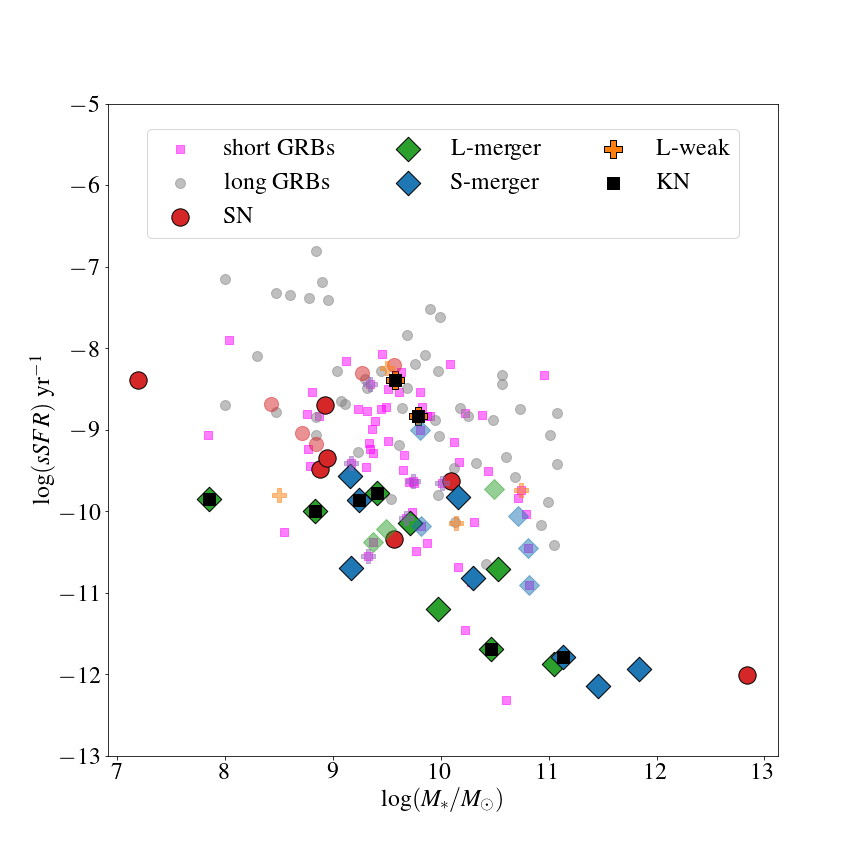}
	\caption{Mass and specific SFRs of GRB host galaxies for long (see Table~\ref{tab:host_props} for references) and short \citep{nugent22} GRBs, overlayed with the sample of {\em Swift} GRBs at $z<0.3$. In general it is apparent that the host galaxies of the L-Merger GRBs are substantially less star forming that those of the Collapsar bursts, and are consistent with those for S-Mergers. We also note the outlier nature of GRB 190829A which is an an exceptionally massive galaxy with a low specific SFR. However, there is apparent star formation close to the location of this GRB (e.g., \citealt{bhi24}).}
	\label{maxmass}
\end{figure}

We next search for trends between the global host galaxy properties of the collapsar and merger-driven GRBs. We obtain host properties for the short GRB hosts from \cite{nugent22} and for long GRBs from the literature (see Table~\ref{tab:host_props}). We have favoured values from sample studies where available, since these are likely to be self-consistent. However, in several cases we report values based on individual burst studies. We note that recomputing and re-fitting all of the available host galaxy data is beyond the scope of this paper, and so some caution should be used in a detailed comparison of individual hosts. However, we expect that the bulk properties can be compared. Where possible we also report the galaxy size, either from the literature, or directly determined from {\em HST} imaging to obtain a 50\% light radius with \texttt{Source Extractor} \citep{Bertin1996}. We catalog the available host properties and galaxy offsets (Tables~\ref{tab:host_props}) and show the stellar mass and SFR relationship within our samples in Figure~\ref{maxmass}.

There are several striking features in this distribution. Firstly, L-merger GRB hosts have substantially lower SFR and higher stellar mass compared to the hosts of most known collapsar and cosmological long GRBs. 
This may be related to the changing locations of star formation with redshift, such that the host galaxy population undergoes substantial evolution. However, importantly, the L-merger sample includes a substantial number of galaxies with very low SFRs such as GRBs 050219, 050724, 060614 and 191019A. In general, similar galaxies to these are not found in the collapsar population, although we do note that the host of GRB 190829A is especially large and massive and as such as a very low specific SFR (the far right point in Figure~\ref{maxmass}). There is evidence for star formation close to the burst site, although the redshift identification for this burst has also been questioned \cite{bhi24}.

\subsection{Offset distribution}

We also consider the offset distribution as a possible proxy for the progenitors (Table~\ref{tab:host_props} and Figure~\ref{gold_sample_box}). Indeed, offsets were considered as strong diagnostics for progenitors before the robust identification of either supernova or kilonovae in cosmological GRBs \citep[e.g.][]{bloom02}. The limited size of our sample precludes  strong constraints regarding the offsets, in particular between L-mergers and the S-mergers. However, the mean offsets are largest for S-Mergers, followed by L-Mergers and lowest for the Collapsars sample, this is in keeping with the expectations for kicked populations. However, it is striking that two L-Mergers have extremely low offsets (GRB 111005A \citep{michalowski18} and GRB 191019A \citep{levan23b}. For GRB 111005A, although the burst is supernova-less to deep limits, its very local origin at 55 Mpc leaves space for alternative progenitor interpretations (see Appendix). For GRB 191019A, the lack of star formation strongly disfavours a massive star origin and perhaps instead indicates a dynamical origin. If the observed merger distribution consists of both long and short GRBs then it may be surprising if two dynamical examples arise in the long population. However, there are also physical rationales for this outcome if, for example, the long duration is due to the interaction of the GRB shock with a dense environment such as an AGN disk \citep{lazzati22,Lazzati_2023}.

\section{Rates} 
\label{sec:rates}
A critical question relating to the populations of GRBs created via different progenitors is the rates of such events. Here we consider the relatively robust type fractions, and also attempt to undertake straightforward rate calculations. We note that full calculations, in which simulations are used to estimate the apperance of a given GRB-lightcurve and its recoverability at different redshifts \citep[e.g.][]{littlejohns13,moss26}, are beyond the scope of this work. 
We discuss several possible selection effects and their potential impact on our rate calculations in Sections~\ref{subsec:mission_bias} and \ref{subsec:sample_sel_bias}. 

\subsection{Type fractions of collapsars and mergers}

The most robust measures that we obtain reflect the observed fractions of different progenitor types. Within the $z<0.3$ sample we calculate type fractions of Collapsars (9 events, 0.31 $\pm$ 0.09), L-mergers (9 events, 0.31 $\pm$ 0.09), S-Mergers (8 events, 0.28 $\pm$ 0.09) and L-Weak (3 events, 0.10 $\pm$ 0.06). The errors on these fractions are determined by the Poisson noise on the relatively small samples. For the $z<0.5$ sample, the fractions are 
Collapsars (15 events $f=$0.25 $\pm$ 0.06), L-mergers (13 events $f=$0.22 $\pm$ 0.06), S-Mergers (13 events, $f=$0.22 $\pm$ 0.06), L-Weak (8 events, $f=$0.13 $\pm$ 0.6) and S-weak (11 events, $f=$, 0.25 $\pm$ 0.06).

We can also ask the related question of the type fractions within the long and short GRB populations. At $z<0.3$ there are 8 short and 21 long-GRBs. Since all the short bursts are in the S-merger class this implies the fraction of short mergers is $f=1.0$, although since the number of trials is limited a more formal estimate is $f>0.7$. 
In the long GRB population, of the 21 bursts, there are \goldlmerger L-mergers ($f=0.43 \pm 0.11$), \goldcollapsar collapsars ($f=0.43 \pm 0.11$) and \goldlweak L-weaks ($f=0.14 \pm 0.08$). The L-weaks could plausibly lie in either class resulting (although the initial numbers of bursts in each is identical). Hence, the total fraction of L-mergers in the long duration class lies in the range $0.32 < f < 0.68$. Since the observed number of collapsars and L-mergers are the same, them under the assumption there are only two progenitor classes, the range for collapsars is identical to that for L-mergers.

\subsection{Sample completeness}
\label{sample_complete}
One of the key questions in understanding the rates of the progenitors inferred from this sample is: \textit{how representative it the sample of the underlying population as a whole?} In particular, only around 25\% of the {\em Swift} GRB sample has known redshifts, implying that our sample is likely not complete at $z<0.5$. While placing strong cuts on population properties (in particular for observability and peak flux) has led to some very complete samples such as BAT6 \citep{Salvaterra2012,DAvanzo2014}, TOUGH \citep{Hjorth2012} and SHOALS \citep{Perley2016}, such approaches at low redshift would reduce already small populations. 

We therefore attempt to answer a different question; namely, \textit{what fraction of $z<0.5$ GRBs are plausibly missing from our population?} To do this we make use of the Legacy Survey Data Release 9 which spans a substantial fraction of the sky and reaches limiting magnitudes ($r = 24$~AB mag; \citealt{Dey2018a}) sufficient to detect most galaxies in our redshift range. For instance, $r=24$ corresponds to $M_B = -17.8$ at $z=0.5$, a factor of two fainter than the Large Magellanic Cloud. Since the majority of {\em Swift} GRBs have X-ray afterglows (although not in ways that are uniform across all durations) we use this as our input catalog of positions. After removing objects marked as point sources, we calculate the $P_{cc}$ (see \cite{bloom02,fong22}) using the positional uncertainty and $r$-band magnitude for all sources within 1~arcminute of the XRT position. We select the source with the lowest $P_{cc}$ as the fiducial ``host'' and retrieve the available spectroscopic redshift or, if unavailable, photometric redshift and its 68\% confidence interval \citep{Zhou+21}. To obtain a sample of ``random'' positions for comparison, we repeat this task 10 times, with the XRT positions offset in random directions by one arcminute. Although the sky coverage of the Legacy Survey DR9 is extensive compared to existing surveys, sufficient data were only available for 46\% of XRT positions. 

\begin{figure}
	\centering
		\includegraphics[width=1.1\columnwidth]{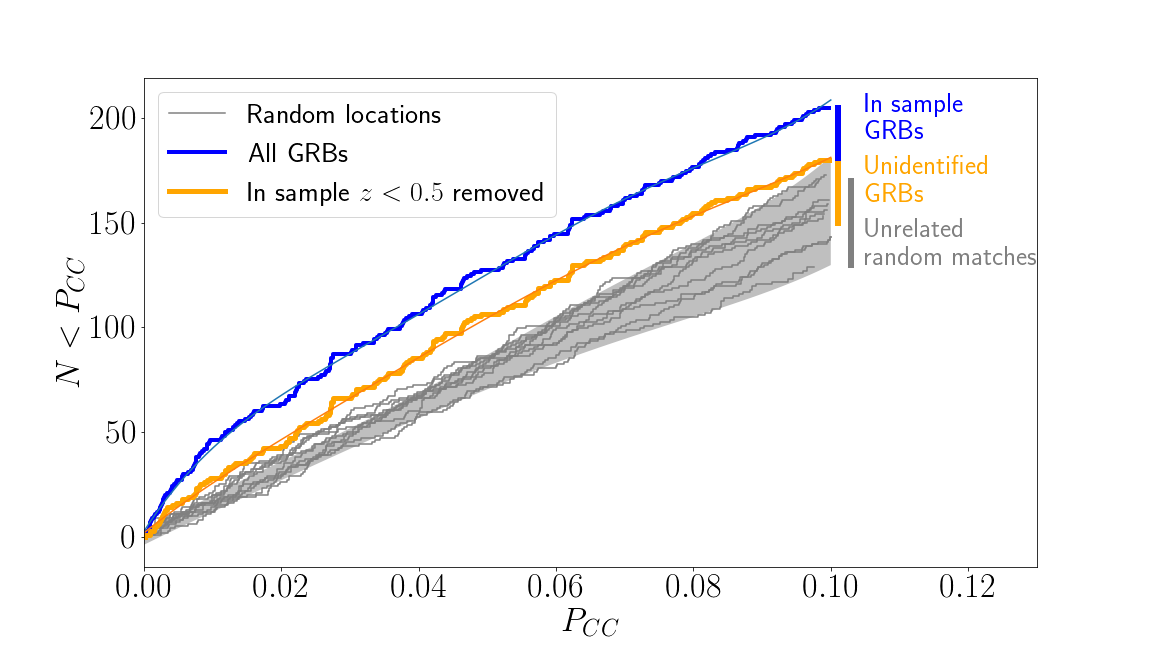}
	\caption{The distribution of chance alignment probability of galaxies at $z<0.5$ in the legacy survey compared to {\em Swift}-XRT positions. The blue line shows the distribution of $P_{CC}$ for the complete {\em Swift}-XRT GRB catalog, the orange line shows the same, but with events already within our samples removed, while the grey lines represent cases where the XRT positions have been offset in random directions by 1 arcminute to provide a set of ``unrelated" burst locations for comparison. It is clear there are a larger number of galaxies in both GRB sets than expected by chance, suggesting there is a substantial, but uncertain population of unrecognised bursts at $z<0.5$ within the {\em Swift} sample. }
	\label{pcc}
\end{figure}

In Figure~\ref{pcc}, we plot the resulting distribution of chance alignment probabilities. We consider a galaxy a possible match if the lower limit of its photometric redshift is $z<0.5$. As can be seen, the actual XRT locations have a substantial excess of galaxies with low $P_{cc}$ compared to the randomized locations. However, since many of these events are already within our sample, we therefore also plot the distribution when the events in the sample have been removed. 

Although there is a substantial range in the recovered numbers for the random locations, the number of GRBs remains above this for both samples considered, leading to the conclusion that there are bursts missing from our $z<0.5$ sample. The random scatter increases substantially at higher $P_{cc}$, but at $P_{cc} =0.05$ we estimate that in our sample there are a total of between 42-58 GRBs at $z<0.5$ of which between 20-36 have previously been identified. Accounting for the sky coverage correction factor of $1/0.46 = 2.17$ these numbers become between 91-126 and 43-78. Given our {\em Swift} sample contains $60$ GRBs this suggests it may be only 48-66\% complete. 

Notably, the excess in number of GRBs over the random locations is not located entirely at very low $P_{CC}$, likely because hosts have already been identified in these cases. At higher $P_{CC}$ the number of random matches exceeds the number of real ones, suggesting it will be difficult to robustly assign hosts in the absence of optical afterglow identifications and redshifts. The implication of this is that many of the true matches in this regime have substantial offsets from their hosts. This may suggest that the population of L-mergers is substantially less complete than for collapsars, as might also be expected given their fainter optical afterglows (Section~\ref{sec:ag_brightness}).

\subsection{Simple rate estimates}
\label{subsec:rate_est}

While useful for understanding the observed frequency of merger-driven GRBs, our type fractions cannot be directly mapped to the volumetric rates of such events. We must construct samples to calculate volumetric rates for comparison to population models and GW inferred merger rates, which are clearly of importance to understanding progenitors. 

Obtaining the detailed event rate for a sample is a difficult problem given the extremely complicated triggering mechanism on {\em Swift} in which the instrument can trigger both in the rate and image domain, and on a wide range of timescales. There are also substantial astrophysical selection effects which may impact any rate estimate, and we describe these briefly below. However, simple rate estimates can still provide potentially valuable information, and we provide these via several routes below, noting the caveats where necessary. 

The volumetric rate is given by

\begin{equation} 
R_{GRB} =  \left({1 \over \Omega \epsilon \eta_{z} } \right) \left({1 \over f_{b} } \right) \left({1 \over Vt} \right)_{z_{max}}.
\end{equation}

Here, the first term describes the contributions due to the fundamentals of GRB observations with {\em Swift}: the fraction of sky surveyed ($\Omega$=0.16, for {\em Swift}) and the duty cycle of the instrument itself ($\epsilon$, which we set to 0.75). The term $\eta_z$ reflects the redshift completeness of the GRB population that we use. For $\eta_z =1$ we would have a complete redshift distribution. Of the $\sim 1600$ bursts detected by {\em Swift} only $\sim 25$\% have redshift measurements. However, one might expect that local GRBs would be more readily identified than the more distant ones, but this is not necessarily the case if they are ejected far from host galaxies, intrinsically faint, or suffer from high host extinction. For simplicity, we use $\eta=1$ in the rate calculations that follow, but note that these could readily be updated to be larger by a factor $\sim 2$ for the incompleteness discussed in Section~\ref{sample_complete}.  
The beaming correction, $f_b$ is highly uncertain but can be constrained via measurements of jet-breaks \citep[e.g.][]{Rouco+23}. Alternatively, setting $f_b$=1 provides only an estimate of the on-axis event rate, which we elect to do in this work. The final terms refer to the survey duration ($t=20$ years for {\em Swift}), and the maximum volume over which the source may be visible. Since cosmological time dilation impacts the effective time at each redshift, we integrate the effective product of survey time and volume from $z=0$ to $z=z_{max}$ to obtain a weighted volume multiplied by time. Since we are interested in creating a volume limited sample to some maximum redshift, we set $V_{max}$ to be either the maximum volume over which a burst was observable, or our redshift threshold ($z=0.3$ or $z=0.5$). Out to $z \sim 0.5$ these effects are modest in magnitude, and indeed could be offset by rate evolution as one moves to higher redshift since both mergers and appear to increase in rate rapidly with increasing redshift \citep{Ghirlanda2022,salafia23}. 

As can be seen, many of these parameters have significant uncertainty which plague the ability to determine robust rates. However, for simplicity, we consider the rate of on-axis events only (i.e. $f_b=1$) with measured redshift ($\eta_z =1$), although we consider how good an approximation this is in Section~\ref{sec:discussion}. In this calculation, the rate is effectively determined by $V_{max}$. 

As bursts are moved to larger distances, their detectability is complex to calculate. The {\em Swift}-BAT has a very large number of possible trigger mechanisms (Section ~\ref{subsec:mission_bias}; \citealt{burns16}). 
At the same time, as a burst shifts in redshift it is also time-dilated and has its spectrum shifted. Hence, fully modelling the complexities of the {\em Swift} response is an extremely difficult task \citep[see also][]{littlejohns13,moss26}. 

As an alternative, it is common to simply use some flux (or fluence) limit and then apply this as the limiting value beyond which a burst cannot be observed. This can then be readily converted to a distance and hence a value for $V_{max}$. This is clearly imprecise as there is no single value at which bursts cease to be detectable, but instead a gradual fall-off in sensitivity that results in far fewer bursts with low peak flux or fluence that expected. Indeed, GRBs approximately follow the -3/2 slope in the $\log({N})-\log(S)$ relation that is expected for a uniform distribution of sources in a Euclidean Universe (although the closeness to this is actually something of a coincidence \citep[see][]{malesani23}). Hence, setting a threshold too high results in many {\em detected} bursts within our samples being formally detected at redshifts greater than they should be detected (i.e. they are fainter than our thresholds). Alternatively, setting them too low underestimates the true rates because many bursts with comparable brightness are missed by the detectors.

To roughly cover the likely possibilities we consider four different cuts here, and our motivation.
The first is a signal-to-noise cut. Since {\em Swift}-BAT has a threshold at 7$\sigma$ we can calculate the difference between the observed signal to noise of the trigger and 7 to estimate the maximum redshift. This tells us out to what distance this particular trigger would have been detectable. However, this forces an underestimation of $V_{max}$ values since a burst may fail to reach one trigger threshold (e.g. a 1 second rate  trigger), but subsequently trigger another (e.g. a 64 second rate trigger). The second approach is to use the observed fluence of the burst -- this approach is more appropriate for image triggers, which do not reach high peak flux levels, and we used $S_{15-150}=2.5 \times 10^{-7}$ erg cm$^{-2}$. A third approach is to use the peak flux in the same way, seeing $F=5 \times 10^{-8}$ erg cm$^{-2}$ s$^{-1}$, and a final version is to use the peak photon flux, setting $P=0.5$ photons cm$^{-2}$ s$^{-1}$. For bursts which are fainter than these values, we continue to consider them, but allow their maximum detectable distances to be less than their observed distance. While this may appear counter-intuitive, it is somewhat analogous to setting a lower threshold, but then including a correction for source incompleteness at that flux level. 
Finally, since most GRB luminosity functions rise strongly to fainter luminosity \citep[e.g.][]{Ghirlanda2022}, the inferred rates are a strong function of the minimum luminosity, with fainter bursts being intrinsically more common. We therefore present rate calculations only for events with $L_{iso} > 10^{48}$ erg, a cut which removes the most local objects (e.g. GRB 111005A) from our samples. 

\begin{figure}
	\centering
		\includegraphics[width=\columnwidth]{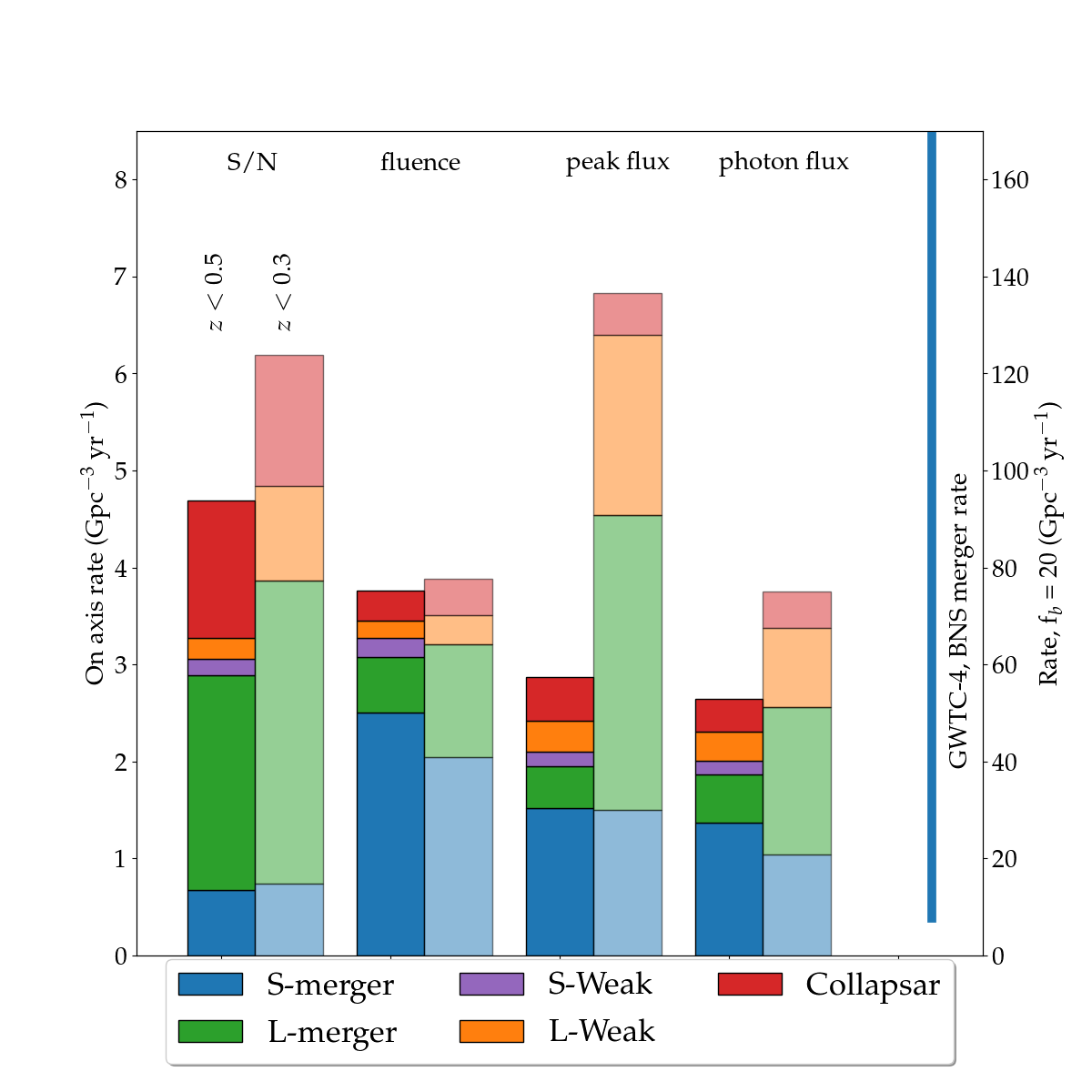}
	\caption{Volumetric rates of our {\em Swift} GRB classes ($z < 0.5$ and $z < 0.3$ is shown in opaque and moderately transparent bars, respectively) calculated using different approaches (Section~\ref{sec:rates}). The bulk differences between the techniques of a factor $\sim 2$ reflect substantial underlying uncertainty but provide general consistency in the relative rates of each progenitor channel. }
	\label{rates}
\end{figure}

The rates resulting from these calculations are shown in Figure~\ref{rates}, and tabulated in Table~\ref{tab:rates}. Despite the differences in these approaches, the rates estimated via these different methods generally agree to each other to within a factor of two, although there are a few cases where larger variance can be seen. This is particularly the case in moving from the peak flux to fluence based approaches, where, for example the L-mergers are signficantly more fluent than the S-mergers, resulting in higher $V_{max}$ (and lower rate)for fluence based calculations in these cases. 

In broad terms, the on-axis rates for Collapsars, L-Mergers, S-Mergers, L-weak and S-Weak bursts are all comparable at the 0.5-2.5 Gpc$^{-1}$ yr$^{-1}$ level for on axis events. Whether this corresponds to comparable actual rate densities for the events depends on if $\eta_z$ and $f_b$ are the same for the different populations, as well as if their luminosity functions behave in similar ways (i.e. the impact of extrapolating the luminosity function below $L_X = 10^{48}$ erg s$^{-1}$, lowering this number would result in far higher rates for collapsar events). If this is not the case, then the true volumetric rates may be quite different to these. In any case, since both of these terms result in upward revision, it is clear that the observed populations can at best provide only lower limits on the true event rates. 

\begin{table}
    \centering
    \caption{Rates (in Gpc$^{-3}$ yr$^{-1}$) of the populations studied here (Section~\ref{sec:methods}) and calculated using different techniques (Section~\ref{sec:rates}). The rates are for on-axis events ($f_b=1$) and on the assumption of 100\% completeness for redshifts $\eta_z =1.0$. The use of different estimators is intended to provide an indication of the uncertainty in the rates, which is clearly substantial. }
    \begin{tabular}{c|cccc}
    \hline \hline  \\
           Swift, $z<0.3$  & S/N & fluence & peak flux & photon flux \\
        \hline
        Collapsar &   1.35 & 0.37 & 0.44 & 0.38\\
        L-Merger & 2.26 & 0.76 & 0.62 & 0.65  \\
        L-Weak &  0.12 & 0.12 & 0.13 & 0.13 \\
        S-Merger &  0.74 & 2.05 & 1.50 & 1.04  \\
        \hline
        Swift $z<0.5$ \\
        \hline 
        Collapsar & 1.41 & 0.32 & 0.45 & 0.34  \\
        L-Merger  & 2.21 & 0.57 & 0.43 & 0.50 \\
        L-Weak &  0.22 & 0.18 & 0.31 & 0.30 \\
        S-Merger & 0.68 & 2.51 & 1.52 & 1.37  \\
        S-Weak & 0.17 & 0.19 & 0.15 & 0.14 \\
        \hline
    \end{tabular}
    \label{tab:rates}
\end{table}

\subsection{Biases due to mission choice}
\label{subsec:mission_bias}

So far, we have confined our analysis to statements about the {\em observed} population seen by {\em Swift}. These may not necessarily be representative of the underlying population of merger and collapsar GRBs due to differences in the triggering process and, to a lesser extent, the softer sensitivity of {\em Swift}-BAT \citep{burns16}. This can be seen in the ratio of short- to long- GRBs observed by different missions. For instance, for BATSE, {\em Fermi}, and {\em Swift}  the fraction of GRBs with $T_{90} < 2$s is 26\%, 21\%, and 9\%, respectively \citep{kouveliotou93,bhat16,lien16}. The implications for the long-GRBs which arise from mergers are less clear, in part due to the difficulty of identifying longer wavelength counterparts to bursts detected with BATSE and {\em Fermi}. 

For the brightest 10 {\em Fermi} bursts, which are more likely to enable IPN localizations, \cite{levan24a} show that at least 2 are associated with kilonovae (GRBs 211211A and GRB 230307A), and further examples remain ambiguous. In the hardness-duration plane these mergers appear to occupy a similar location to the bulk of the long GRB population although several lie at the softer end of the distribution (Figure~\ref{hardness}). In summary, we conclude that it is plausible but uncertain that a comparable fraction of L-Mergers are detected by {\em Fermi} and {\em Swift}, though a larger sample is needed to confirm these results.

Extending further to our sample which contains all bursts at $z<0.5$, adds a further \totbronze GRBs to consider (\totallcats total). The additions as a result of this sample selection are predominantly of bursts detected prior to the launch of {\em Swift} in an era without rapid positions and where afterglows needed to be searched for. They are also from detectors optimised to even softer energies than the {\em Swift}-BAT (e.g. the BeppoSAX WFC which reached down to 2 keV, or the HETE-2 WXC which also went to 2 keV). A number of bursts are detected with the {\em Fermi}-Gamma-ray Burst Monitor (GBM; \citealt{meegan09}), which has much harder sensitivity. More recently the Einstein Probe has delivered substantial samples of Fast X-ray Transients whose progenitors appear related to GRBs, but we do not consider directly in our rate calculations. We discuss this further in Section~\ref{sec:discuss_othertransients}. 

The differing between the type fractions of the {\rm Swift} and all mission samples (Table~\ref{tab:numbers_by_cat}) may indicate that selection effects are impacting detection rates. In particular, if the relative faintness of L-Merger afterglows is a general feature of the population, then these may have been missed in an era where rapid follow-up was much rarer. Alternatively, it could reflect the small number statistics nature of these analyses, which, even with samples of tens of bursts result in the type fractions being relatively poorly constrained. 

\subsection{Biases due to sample selection} 
\label{subsec:sample_sel_bias}

 One potential source of uncertainty in this sample is identification of the burst's redshift. In particular, the redshifts recorded for short GRBs are often based on the most likely galaxy and the $P_{cc}$ method (e.g., \citealt{fong22}) which implies that a small fraction of associations might be spurious. Further, the large kick velocities expected for some BNS mergers \citep{Tauris+17} and resulting large host galaxy offsets ($\sim$10-100~kpc; \citealt{fong13,fong22}) indicate that some nearby mergers might be missed. 
 
Secondly, within the long GRB population X-ray afterglows are near ubiquitous. However, in the short GRBs X-ray afterglows are only recovered in $\sim 70$\% of cases \citep{fong15}. It is unclear if the afterglow faintness is related to distance (i.e. bursts without afterglows are more distant than our sample redshift cut), or intrinsic faintness, but the incompleteness of the short GRB afterglow catalog means it is plausible that further nearby events are missing.  Indeed, several short GRBs lack X-ray or optical afterglows within their error boxes, but do contain relatively nearby bright galaxies, e.g. GRB 050906 \citep{levan08} and GRB 090417A \citep{fox09}. This may mean the fraction of recovered afterglows, and hence of identified redshifts and host galaxies, is not constant between different progenitor populations. 

Furthermore, our rate estimates are based on the detectabilty of the prompt emission, while in practice to be included in the sample bursts must be localised and a redshift obtained, either from the afterglow or the host galaxy. This may not always be trivial. Several L-mergers are extremely bright, such that they would be visible to anywhere within the horizon considered for the calculations. For example, GRBs 230307A and 211211A are the second and seventh brightest bursts seen by the {\em Fermi}/GBM \citep{Burns2023,levan23a}. However, while the prompt emission is extremely bright, the afterglows and host galaxies are not. In both cases, at $z=0.5$ the afterglow magnitudes at the time of discovery would have been $r>24$ if at $z=0.5$. Given that neither event lies directly on its putative host galaxy it is likely that neither of these events would have been localised or recognised at $z=0.5$, despite both being detectable at this redshift in $\gamma$-rays. The relatively faint optical afterglows for these events suggest that they would not have been identified far beyond their observed distances. Since the difference in co-moving volume between $z=0.08$ and $z=0.5$ is a factor of $>250$ the implications for the rate of discovery of L-mergers is substantial. 

\section{Discussion}
\label{sec:discussion}

\subsection{Implications of a substantial population of mergers in the long GRB population}
\label{subsec:implications}

The core implication of this work is that in the local ($z < 0.5$) Universe the observed fraction of GRBs from collapsars that form supernovae and progenitors which do not form supernovae is similar. For our sample, the differences in the afterglow brightness, the detection of kilonova in some cases, and the presence of the progenitors in apparently ancient populations all point to a different progenitor, in at least the majority of the supernova-less GRBs. The simplest explanation (i.e. the one which does not require an additional progenitor channel) is therefore that they arise from compact object mergers. Indeed, their environments, offsets and hydrogen column densities are all consistent with the short-GRB population (Section~\ref{sec:core_props}).

However, in principle, there are several alternative explanations for the non-detection of supernovae in GRB afterglows. For example, the initial collapsar model of 
\cite{woosley93} was developed to explain failed supernovae, in which much material is accreted onto the nascent compact object and a jet is launched, but little or no material is ejected as a supernova. If the process of stellar collapse in GRBs produces a dichotomy between events that launch nickel driven supernovae, and events that do not (Figure~\ref{fig:nickelandlum}), then it is plausible that L-Mergers could be linked to massive stars \citep[e.g.][]{fynbo06}. Hence, while evidence suggests that many SN-less long GRBs are mergers,  we should continue to be cautious about interpreting all events as arising via this mechanism. Indeed, some rare events appear to defy classification into the existing progenitor schemes, implying some rare channels may also operate \citep[e.g.][]{levan14,levan25b,neights25,gompertz25}. This is discussed further within our follow-up companion paper (Rastinejad et al. in prep.).

At least within our local sample, there is no evidence for overlap of supernova progenitors in otherwise short GRBs. This is in contrast to the expectations of some models \citep[e.g][]{bromberg13}, although a more recent analysis of their sample of observations implies that at low redshift mergers may well dominate, supporting the conclusions of this paper \citep{lr26}.

There are some challenges to this simple picture. For example, the long GRB 111005A lies in a galaxy at only 55 Mpc. Follow-up was very limited due to the proximity of this source to the Sun. However, deep limits on supernova emission were possible thanks to {\em Spitzer} observations \citep{michalowski18}. Strikingly, because it was so nearby,
these are also strong limits on any kilonova emission (see Appendix \ref{appendix:GRB 111005A}). It is possible, therefore, that GRB 111005A arises from a different progenitor. Given the very different energy scale, this is not an unreasonable proposition, although it should be noted that it has a similar energy to GRB\,980425/SN~1998bw and GRB\,170817A/AT\,2017gfo -- the proto-types of GRB supernovae and kilonovae respectively. Alternatively, the associated kilonova must be much fainter than those previously seen. This is also plausible; the expected peak luminosities of kilonovae extend to much fainter levels than seen to date (e.g., \citealt{Metzger19}). Indeed, ejecta masses inferred for kilonovae show hints of diversity \citep{Rastinejad+25} and the high median ejecta masses are problematic compared to expectations of many merger simulations \citep[e.g.][]{Siegel19}, though this may also be due to uncertainties in modeling \citep[e.g.][]{SarinRosswog24}. 

\subsection{Redshift evolution}
The ratio of collapsar to merger GRBs is expected to be a strong function of the delay time distribution for compact object mergers and the star formation history of the stellar population from which the binary emerged. Collapsars occur promptly with star formation, while mergers follow some delay time distribution. If observing a single burst of star formation, the ratio of collapsars to mergers would evolve strongly from collapsar to merger dominated as the population aged. 

Cosmologically, the order of magnitude increase in the Universal star formation rate between the current time and $z \sim 1$ also results in an increase in the collapsar GRB rate. However, this increase is even larger due to the metallicity bias in collapsar GRB production \citep[e.g.][]{fruchter06,graham13,perley16b,disberg25}. Indeed, in the model of \cite{Ghirlanda2022} (fit to all long GRB data) the rate increases as $(1+z)^{3.2}$ out to $z \sim 3$. 

Compact object binary production (i.e. the initial production of the NS-NS or NS-BH system) requires two supernovae and so also follows the star formation rate. However, there is then a delay time distribution due to the gravitational wave inspiral time. It common to approximate the delay time distribution as $t^{-1}$ \citep{piran92}, whereby there are approximately equal number of mergers per decade in time, after some initial delay.

\begin{figure}
	\centering
		\includegraphics[width=\columnwidth]{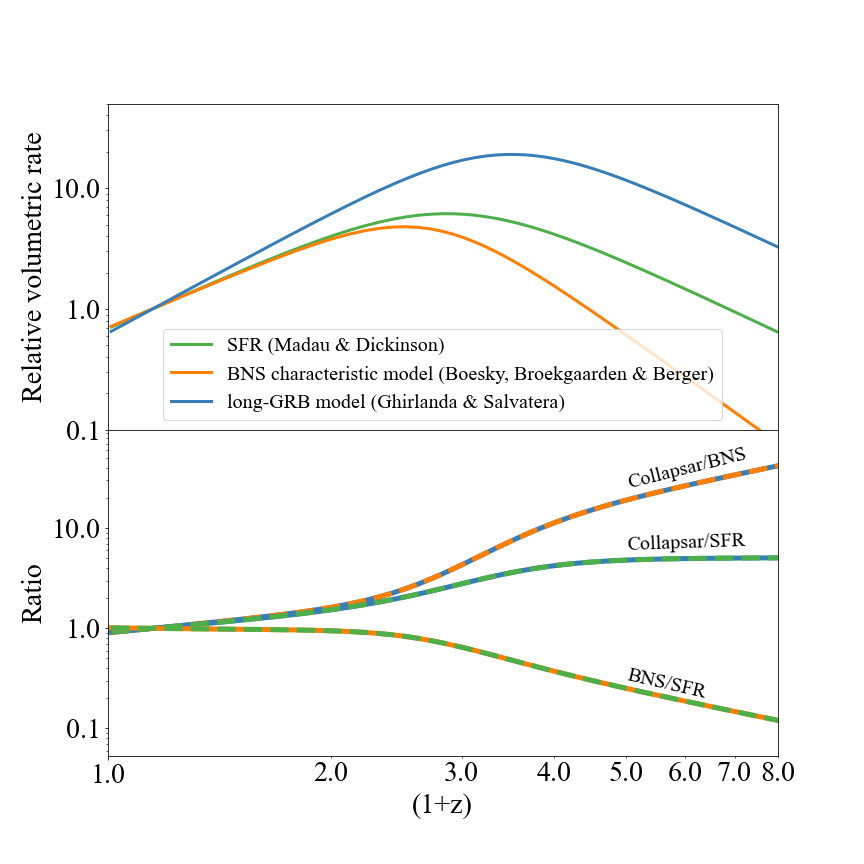}
	\caption{Characteristic cosmic evolution in the populations of BNSs and collapsars. In general, collapsars are expected to form preferentially in low metallicity environments and so occur on average at higher redshifts than the bulk of the star formation. In contrast, BNS mergers form via gravitational radiation induced orbital shrinkage which follows the SFR by some delay time. As a consequence of this, if similar numbers of collapsars and BNS mergers exist in the local Universe, the numbers at high redshift (beyond $z>2$) will be overwhelmingly dominated by the collapsar population.}
	\label{fig:ratio_evolution}
\end{figure}

Hence, we might expect that at low redshift where the star formation rate is lowest, the ratio of collapsars to mergers will be lowest, while it will become higher at larger redshift. A full treatment of the possibilities is beyond the scope of this paper, but an outline of this scenario is shown in Figure~\ref{fig:ratio_evolution}, where we plot the cosmic SFR, the long GRB rate and a characteristic BNS merger rate (estimated from \cite{boesky24}, although we note that the true BNS merger rate remains highly uncertain and many different paramterisations exist \citep[e.g.][]{salafia23}). These rates are noramlised to be equal at $z=0.1$ so that the evolution can be seen (i.e. we plot relative rather than absolute rates). In particular, because of the metallicity bias for collapsars and the delay time distribution for mergers, the ratio of collapsars to the SFR increases with time while the ratio of mergers decreases. The result is that relative rates of collapsars to mergers strongly favours collapsars beyond $z \sim 1$, and even changes by a factor of two out to $z=1$. 

Coupled to these intrinsic effects are the selection effects surrounding afterglows, as shown in Section~\ref{sec:ag_brightness}, the afterglows of merger-driven GRBs are typically fainter than for collapsars. We therefore suggest that optical selection may effectively enforce a collapsar dominated sample, especially at higher redshift. As can be seen in Figure~\ref{fig:xray_opt_aglcs}, at $z=0.5$, none of the merger driven GRBs are brighter than $R\sim 22$ on timescales beyond a few hours. Hence, obtaining accurate positions and afterglow spectroscopy for these events would be extremely challenging. Indeed, the prevalence of collapsars in pre-{\em Swift} samples may readily be explained by such an effect, since afterglow observations were typically undertaken at later times, and may only have found brighter afterglows. 

\subsection{Relation to other high energy transients}
\label{sec:discuss_othertransients}
In this work we have confined our analysis to populations detected by GRB detecting missions. However, this boundary is not always clear, and it has long been recognised that GRBs with much lower peak energies exist, often into the X-ray regime. Indeed, some of our sources (e.g. GRB\,060218) have previously been given the designations of X-ray Flashes (XRFs) where the X-rays ($E_{\rm peak} \lesssim 15$~keV) dominate the emission \citep[e.g.][]{soderberg04}. In addition, there is a population of high energy transient detected only in X-rays, the FXTs that have been localised by focusing X-ray telescopes in the past 20 years \citep[e.g.][]{jonker13,alp20,qv22,qv23} and whose origin has been subject to substantial debate, with mergers often suggested as a favoured model \citep[e.g.][]{Xue+19}. This field has been transformed in the past 2 years following the launch of the Einstein Probe \citep{Yuan+22} that has identified over 200 FXTs with arcminute-scale positions and provided redshift measurements for $>50$ events \citep[e.g.][]{oconnor25}. Around 30\% of these FXTs have been associated with GRB events, mostly identified by {\em Fermi} (Ravasio et al. submitted) and would in principle meet our sample definition, while the remainder do not. 

Of the FXT population detected to date there are 6 EP events reported to the GCN with $z<0.5$. Of the triggered events these are EP240414a ($z=0.40$; \citealt{Li2025}), EP240506a \citep[$z=0.12$][]{Liang2026}, EP 250108a ($z=0.18$; \citealt{EylesFerris2025,Rastinejad+25_ep,Li2025,Srinivasaragavan+25}), EP250304a ($z=0.20$; \citealt{Cotter+26}),  EP260321a ($z=0.0334$; \citealt{amc26,chen26,oconnor26,Rastinejad+26,yuan26}), while one, EP250827b ($z=0.12$; \citealt{Srinivasaragavan+25_ep250827b}) was identified based on an optical association. We note that two further events have been suggested to be a low redshift, and no supernova has been reported, sometimes to deep limits. These events, EP241107a \citep{eap26} and EP250207b \citep{jonker26,beccera25} could be prime candidates to be associated with mergers. However, more recent observations suggest both are actually at much higher redshift (Goodson et al. in prep, van Hoof et al., in prep). Hence, the majority, if not all of the EP detected low-z events are associated with broad-lined type Ic supernovae, and appear to hence be related to the collapsar GRBs (e.g., \citealt{vandalen25,Rastinejad+25_ep,Srinivasaragavan+25,Srinivasaragavan+25_ep250827b, Cotter+26}). However, none of these are associated with GRB emission implying that they may well represent the soft, low luminosity end of the population. 

There have been two FXTs detected coincident with short-GRBs, EP250704a/GRB250704a \citep{lian26} and EP260527a/GRB260527A \citep{svinkin26,yang26}. However, both of these are at redshifts beyond our sample cut. Hence, the within our sample selection the EP population would appear dominated by collapsar events. Their inclusion would hence substantially alter our conclusions, suggesting a domination of collapsar-like events. However, the differing selection function means these conclusions cannot be readily extrapolated to the $\gamma$-ray detected population. 

Furthermore, there are several other possible transient classes that have been suggested to create $\gamma$-ray emission, including the tidal disruption of either main sequence  white dwarf stars around massive black holes (or intermediate mass black holes in the case of white dwarfs), or unusual stellar mergers. These events are typically invoked to explain events with much long durations, the so-called ultra-long GRBs \citep{levan14}, which may themselves be a heterogeneous population \citep[e.g.][]{Greiner2015,levan25b,gompertz25,neights25,oconnor26}. However, it is also possible that alternative progenitor populations could exist within our sample. Perhaps prominent amongst bursts for which this may be the case is GRB 111005A (Section~\ref{subsec:implications}), for which an interpretation with a different progenitor is perhaps most readily made in this case. 

Finally, the class of Luminous Fast Blue Optical Transients (LFBOTs, e.g., \citealt{prentice18,perley19,ho23}) are also characterised by high energy emission extending at least into the X-ray regime with luminosity typically a factor of 10-100 fainter than luminous GRB afterglows, but of rather similar luminosity to the mergers and low luminosity collapsars \citep[e.g.][]{chrimes24}. These sources have not be observed at the peak of their X-ray lightcurves, and it is unclear if they could also create $\gamma$-ray emission, although they are at least moderately relativistic explosions. The LFBOT population is poorly defined, but a feature of the most luminous and bluest examples is that they decay rapidly without any substantial contribution from nickel, and do not transition into any readily identifiable supernova sub-type. Should any of these progenitors exist within our sample it is possible they could skew our merger fraction to be larger than it otherwise would be. Perhaps the most notable example amongst our GRBs in this regard is GRB 060614 whose lightcurve exhibits a sharp rise to a peak at $\sim 0.2$ days with an absolute magnitude of $\sim -20$, followed by a rapid decay \citep{mangano07}. However, the X-rays exhibit a somewhat similar morphology in this case, and so an afterglow origin with energy injection is also entirely plausible. The suggested presence of a kilonova would favour a merger origin in this case of this burst (e.g., \citealt{jin15}).

\subsection{GW - GRB coincidence}
\label{subsec:gwgrb}

The near simultaneous detection of a GRB and a GW chirp provides a prompt, narrower localization enabling an easier route to the identification of the electromagnetic counterparts to GW sources. Furthermore, the precise time window greatly narrows down the effective number of trials for GW searches and results in an increased horizon such that sub-threshold events can be identified if a GRB is also observed \citep[see e.g.][]{sarin23}. Although GRB emission is likely only detectable for the fraction of mergers which are detected close to face on, this is also the orientation at which the GW strain is maximized so that crudely we expect to see substantially more mergers face on that would be expected for a truly isotropic selection.  However, even $z \sim 0.3$ is substantially beyond the threshold for binary neutron detection with the current generation of interferometers. 

Strikingly, this situation is likely to be reversed for the next generation of GW detectors, Einstein Telescope (ET) and Cosmic Explorer (CE), in which the horizon distances are an order of magnitude larger than the current interferometer networks at design sensitivity. For these detectors, the detection of (near) isotropic signals such as kilonovae is likely to be substantially more arduous, and will require facilities such as the NSF-DOE Vera Rubin Observatory to obtain the requisite depths \citep{andreoni22}, and even in these cases only be sensitive to mergers in the relatively local Universe. An AT2017gfo-like kilonova peaks at $r \sim 25$ at $z=0.3$, and many $r \sim 25$ transients may be present in a typical GW localization. On the other hand, both long and short GRBs will be more readily identified. Because of the increase in the search volumes, the rates of mergers detected will be larger than that seen by 2G detectors by a factor of the volume probed (along with any cosmological rate variation), and should identify several thousand BNS mergers per year. We may be able to rely on the samples which are face-on and associated with GRBs \citep{bluebook}. Our results suggest that long GRBs may well be a powerful route in this era. However for this to work in practice it will be necessary to have a sufficiently sensitive $\gamma$-ray detector. Indeed, the population of merger GRBs (so far considered predominantly to be those of short-duration) appears to have a lower median redshift compared to the long GRBs \citep{jakobsson06,nugent22}. Given that the horizons of ET and CE for binary neutron star mergers extend to $z>3$ for neutron star mergers, the ideal $\gamma$-ray systems would have sensitivities substantially higher than {\em Fermi} and {\em Swift}, while still providing wide-area coverage and ideally good enough localization for narrow field X-ray and optical/IR instrumentation such as Athena and Extremely Large Telescopes.

We finally note that GW-GRB coincidence may ultimately be the most robust route to distinguish between the different progenitor channels, in particularly those which involve the mergers of white dwarfs, neutron stars and black holes. GW detections can directly determine the masses involved for NS-NS and NS-BH systems, providing robust indications of the progenitor systems. Mergers involving white dwarfs emit gravitational waves at much lower frequency and the non-detection in high frequency interferometers (e.g. LVK, ET, CE) would favour such a scenario. The lower GW frequency produced by mergers involving white dwarfs means that they will not be detectable to ground-based interferometers, although if some GRBs are related to such mergers then they would be in detectable to proposed spaced-based detectors such as DECIGO \citep[e.g.][]{yin23}.

\section{Conclusions}
\label{sec:conclusion}
We have presented the properties of a sample of GRBs identified by the {\em Swift} satellite and other missions at $z<0.5$. Our main conclusions are as follows:

\begin{enumerate}
    \item Strikingly, in contrast to the generally accepted picture of GRB progenitors in which long GRBs arise in core collapse supernovae and short GRBs in compact object mergers, we find that the number of long GRBs for which there is no supernovae (L-Mergers) is comparable in number to the sample for which a supernova is identified (Collapsars). 
    \item We demonstrate that, in general, the non-detection of supernovae is to levels substantially fainter than the bulk of the supernova luminosity function and no events are consistent with the peak of the core collapse supernova distribution at $M_V \sim -17$. This lack of supernova cannot be attributed to a broad spread in supernova luminosities. 
    \item We find the fraction of L-Mergers in our $z<0.3$ sample is comparable to that of short GRBs from mergers (S-Mergers), suggesting that a substantial fraction of $z<0.5$ mergers produce long GRBs. 
    \item We show that the bulk properties (afterglow brightness, X-ray absorption, host galaxy properties) of the bursts without identified supernovae are offset from those of the supernova related bursts (although given the small number of bursts in sample this is not of high statistical significance) in the directions of fainter afterglows, lower X-ray absorption and less star forming host galaxies. Combined with the presence of kilonovae in several of the supernova-less examples we suggest that the most economical suggestion is that these bursts arise from compact object mergers. 
\end{enumerate} 

If correct, our results suggest that {\em in the local Universe} the {\em observed} number of long and short-GRBs from compact object mergers are comparable, and implies that long-GRBs from mergers are in fact a frequent, rather than rare outcome. That more distant GRB populations are dominated by collapsar GRBs is expected due to the strong evolution in the cosmic SFR. However, this outcome implies that our simplistic progenitor outcomes are in need of substantial revision.

\section*{Acknowledgements}

JCR was supported by NASA through the NASA Hubble Fellowship grant \#HST-HF2-51587.001-A awarded by the Space Telescope Science Institute, which is operated by the Association of Universities for Research in Astronomy, Inc., for NASA, under contract NAS5-26555.  IM acknowledges support from the Australian Research Council (ARC) center of Excellence for Gravitational Wave Discovery (OzGrav), through project number CE230100016.

\section*{Data Availability}

This paper does not use substantial ``new" data, but rather compiles data from various sources as indicated in the text. The various sources of data are indicated, and a summary of these is provided in the relevant tables throughout the article.



\bibliographystyle{mnras}
\bibliography{refs}



\appendix
\onecolumn

\section{Supernova Constraints}
\subsection{Nickel Mass Estimates}

\label{subsec:ni56_limits}
Nickel mass ($M_{Ni}$) and luminosity are proportional to one another through the formula from \cite{2005A&A...431..423S} based on \cite{1982ApJ...253..785A}:
\begin{equation}
    \frac{M_{Ni}}{M_\odot} = L_p  (10^{43} \text{erg} s^{-1})^{-1} (6.45 e^{\frac{-t_p}{8.8}} + 1.45  e^{\frac{-t_p}{111.3}})^{-1},
\end{equation}
\noindent where $L_p$ is peak bolometric luminosity and $t_p$ represents the rise time. We adapt \citet{2016MNRAS.458.2973P} to be a ratio of $M_{Ni}$ and luminosity for SN~1998bw. By setting $t_p$ as a constant, the proportionality relationship can be re-arranged for ($M_{Ni}$) and scaled to 1998bw as: 
\begin{equation}
M_{Ni} =  M_{Ni_{1998bw}} \frac{L_p}{L_{p_{1998bw}}}.
\end{equation}

We note this method is an overestimate of the true $^{56}$Ni masses \citep{Khatami_2019}, however, it is suitable for the comparative analysis of this work. These mass estimates for supernova-less GRBs are shown in Figure \ref{fig:nickelandlum}. 

\subsection{Summary Tables}
We record the observations and derived properties of the bursts used in our analysis in Tables~\ref{sample} (bursts in each sample, core properties and progenitor and duration classes), \ref{sample_photometry} (observations used to derive $f_{\rm 1998bw}$) and \ref{tab:host_props} (host galaxy and afterglow properties).

\input{WRAPPED_TABLE_D1_COMPLETE}

\newpage

\input{F_98s_HERE}

\newpage

\input{WRAPPED_host_prop_table}

\section{Summary of individual burst analysis}
\label{individualGRBs}
We do not provide a burst by burst analysis here, but note that the majority of the comparison data is obtained from the literature with the relevant citations provided in Table~\ref{sample} as well as in the main text for large samples. However, in a small number of cases we have undertaken additional analysis to provide further information, and we summarise this below. 
\newline
\newline
\subsection{GRB 111005A}
\label{appendix:GRB 111005A}
\begin{figure}
	\centering
		\includegraphics[width=0.5\columnwidth]{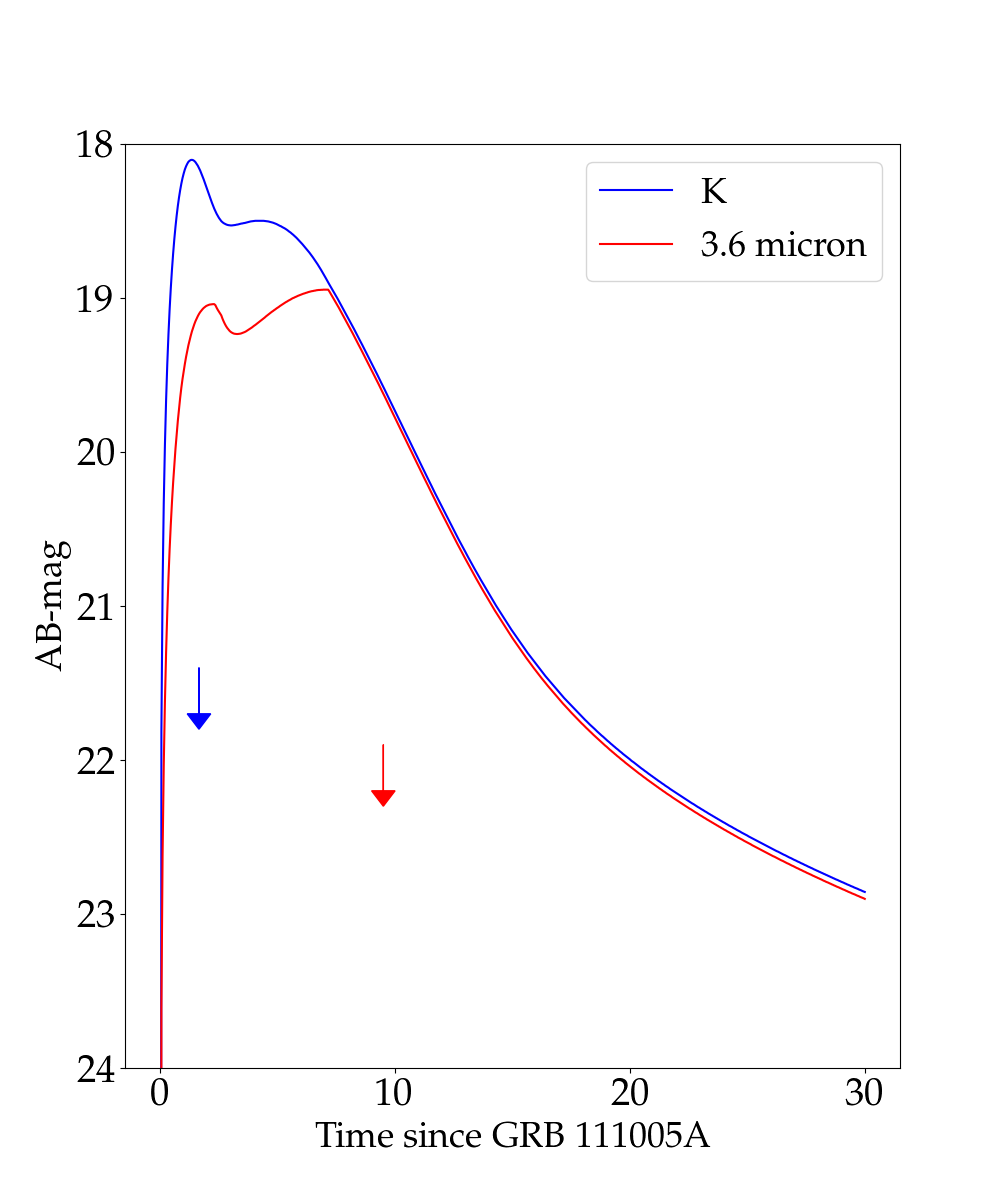}
	\caption{{\em Spitzer} and ground based IR limits on the observed flux of GRB 111005A compared to spectral models of AT2017gfo. The limits correspond to absolute magnitudes in the range -11 to -12, and are substantially fainter than AT2017gfo and the majority of kilonova models. Since these observations are in the mid-IR it is difficult to invoke extinction to explain this difference unless it is extremely large, and this challenges to possibility that GRB 111005A created a kilonova.}
	\label{fig:111005a}
\end{figure}

Our method finds a flux density ratio to SN 1998bw to be $f_{\rm 1998bw}$ <0.149. Despite this, we note the limitation of our method and set this to be $f_{\rm 1998bw}$ < 0.02 based of data from NASA's Spitzer Space Telescope as reported in \cite{michalowski18}. We note this GRB suffers from an uncertain yet significant amount of dust extinction, and does not have an X-ray afterglow. It is notable that it is one of the closest bursts in the sample (after SN1998bw/GRB980425) and has a particularly low luminosity. Indeed, at this redshift, the limits are not only strongly constraining of supernova emission but also of kilonova emission. If GRB 111005A is created by a compact object merger as also suggested by \cite{dong18}, then any associated kilonova is also substantially fainter than AT2017gfo as indicated in Figure~\ref{fig:111005a}. Hence, while we retain this GRB within our L-merger sample, we note that its progenitor is particuarly uncertain. Furthermore, as a very low luminosity event, it does not meet the required luminosity $L> 10^{48}$ erg s$^{-1}$ for inclusion in rate calculations.  \newline\newline

\subsection{GRB 051109B}
For GRB 051109B observations were serendipitously obtained via the CRTS (Catalina Real-Time Transient Survey). We retrieved these images, confirmed that no new transient was present compared to images taken before the burst (on Oct 31) and obtained limits calibrated to the PanSTARRS catalog. We show these limits in Table~\ref{tab:051109B}. Despite the shallow limits, the nearby nature of the burst is such that moderately deep constraints can be put on the presence of a supernova under the assumption of zero host extinction. 

\begin{table}
    \centering
    \caption{CRTS photometry of GRB 051109B. $\Delta T$ refers to time since GRB\,051109B.}
    \begin{tabular}{lll}
    \hline \hline
        Date & $\Delta T$ & Limit \\
         & (days) & (AB mag) \\
        \hline
        2005-11-10 & 0.74 & 20.5 \\
        2005-11-12 & 2.82 & 20.9 \\
        2005-11-20 & 10.76 & 20.4 \\
        2005-12-04 & 24.74 & 20.4 \\
        2005-12-10 & 30.73 & 20.5 \\
        2005-12-30 & 50.77 & 20.7 \\
        \hline \hline
    \end{tabular}
    \label{tab:051109B}
\end{table}

\subsection{GRB 100316D}
In Table \ref{sample} we note two values of $f_{\rm 1998bw}$ due to the two values of host extinction published in the literature. 
We set a lower limit on the SN 1998bw flux density fraction of $f_{\rm 1998bw}$=0.26 based on \cite{starling11}, and $f_{\rm 1998bw}$=1.85 as an upper limit corresponding to \cite{bufano12}.

\subsection{GRB 190829A}

Similar to GRB 100316D, there are multiple values of host extinction in the literature and we account for this in our method. We calculate the lower limit on the SN 1998bw flux density fraction $f_{\rm 1998bw}$=0.6 \cite{Huang_2023}, and and upper limit of $f_{\rm 1998bw}$=1.32 \cite{chand20}. \newline\newline
\subsection{GRB 190114C}
GRB 190114C was an extremely luminous GRB, for which our simple approach to estimating the supernova luminosity fails. In particular, the ratio to SN 1998bw is 3.57 considering the high extinction along the line of sight \cite{Melandri_2022}.  While there is strong evidence for an associated supernova, a measure of the supernova brightness cannot readily be obtained without a detailed afterglow + SN fit. This is also the case for GRB 221009A. 

\subsection{GRB 171010A}
Due to the bright nature of the host galaxy of GRB 171010A, host subtraction was performed using a host absolute magnitude value of -20.8 (r-band) \citep{Melandri+19} in order to obtain an estimate of the afterglow and/or supernova brightness. We note this may introduce additional uncertainty in the reported value of $f_{1998bw}$ for this burst and encourage additional caution when interpreting this value. 


\bsp	
\label{lastpage}
\end{document}

%% file: WRAPPED_TABLE_D1_COMPLETE.tex
\setlength{\tabcolsep}{3pt} 
\captionof{table}{Samples of bursts at $z<0.3$ and $z<0.5$. The core diagnostics of interest are shown including the duration, redshift and presence or absence of a SN or KN (Y=Yes, N=No, A=Ambiguous/no-search). We note that we do not assess the level of evidence for individual SN or KN, but report claims for their presence (or absence) from the literature. For {\em Swift} and {\em Fermi} bursts the duration is taken from the relevant catalogues \citep{lien16,vonkienlin20}.}
\label{sample}

\begin{xltabular}{\textwidth}{l c l l c l c c c X}
    \hline \hline
    GRB & Mission & $T_{90}$ & $z$ & SN? & SN lum. & KN? & Duration Class & Prog. Class & References \\ 
     & & (s) &  &  & ($f_{\rm 1998bw}$) & & & & \\
    \hline \endhead
    \hline
    \multicolumn{10}{r}{\textit{Table B2, Continued on next page}} \\
    \endfoot

    \hline
    \endlastfoot
    \multicolumn{10}{l}{\textbf{\textit{Swift}, $z<0.3$} sample} \\
    \hline
    111005A & Swift & 23.21$\pm$5.06 & 0.013 & N & <0.02 & N & Long & L-Merger & \cite{michalowski18}\\ 
    060218 & Swift & $\sim$ 2100 & 0.0335 & Y & 0.85 & N & Long & Collapsar & \cite{Kocevski+07}\\ 
    171205A & Swift & 190.47$\pm$33.88 & 0.0368 & Y & 0.63 & N & Long & Collapsar &  \cite{izzo19} \\ 
    100316D & Swift & 521.88$\pm$439.62 & 0.0591 & Y & 0.30 - 2.10 & N & Long & Collapsar & \cite{starling11}, \cite{bufano12} \\ 
    190829A & Swift & 56.89$\pm$47.00 & 0.0785 & Y &  0.74 - 1.64 & N & Long & Collapsar & \cite{hu21}, \cite{Huang_2023} \\ 
    211211A & Swift & 50.71$\pm$0.93 & 0.076 & N & <0.01 & Y & Long & L-Merger & \cite{rastinejad22}, \cite{troja22}\\ 
    051109B & Swift & 15.70$\pm$4.12 & 0.08 & N & <0.26 & N & Long & L-Merger & \cite{perley06}, This work \\ 
    060505 & Swift & $\sim 4$ & 0.089 & N & <0.01 & Y & Long & L-Merger & \cite{fynbo06}, \cite{ofek07} \\ 
    080517 & Swift & 64.51$\pm$22.30 & 0.09 & A & - & N & Long & L-Weak &  \cite{stanway15} \\ 
    150101B & Swift & 0.012$\pm$0.009 & 0.13 & N & <0.02 & Y & Short & S-Merger & \cite{fong16}, \cite{fong22}, \cite{troja18} \\ 
    180728A & Swift & 8.68$\pm$0.30 & 0.117 & Y & 0.93 & N & Long & Collapsar & \cite{Rossi+26}  \\ 
    080905A & Swift & 1.02$\pm$0.01 & 0.122 & N & <0.03 & N & Short & S-Merger & \cite{rowlinson10}, \cite{Guelbenzu_2012}\\ 
    060614 & Swift & 109.10$\pm$3.37 & 0.13 & N & <0.02 & Y & Long & L-Merger & \cite{fynbo06}, \cite{gal-yam06}, \cite{gehrels06}, \cite{2009_Xu}, \cite{jin15} \\
    161219B & Swift & 6.93$\pm$0.79 & 0.147 & Y & 1.17 & N & Long & Collapsar & \cite{Cano+17} \\ 
    221009A & Swift & 1104.75$\pm$7.64 & 0.151 & Y & 1.37 & N & Long & Collapsar & \cite{levan23a}, \cite{blanchard23}, \cite{Laskar+23}\\ 
    130822A & Swift & 0.044$\pm$0.011 & 0.154 & N & <0.16 & N & Short & S-Merger &  \cite{fong22}, \cite{rastinejad21}  \\ 
    160821B & Swift & 0.48$\pm$0.07 & 0.162 & N & <0.01 & Y & Short & S-Merger & \cite{lamb19}, \cite{troja19}, \cite{kasliwal17}\\ 
    050219A & Swift & 23.81$\pm$2.26 & 0.211 & A & - & N & Long & L-Merger & \cite{rossi14} \\ 
    050509B & Swift & 0.024$\pm$0.010 & 0.225 & N & <0.04 & N & Short & S-Merger &  \cite{bloom05}, \cite{hjorth05}, \cite{Castro_2005}, \cite{Cenko_2005}\\ 
    120305A & Swift & 0.099$\pm$0.012 & 0.225 & N & - & N & Short & S-Merger & \cite{fong22} \\ 
    211227A & Swift & 83.51$\pm$7.56 & 0.228 & N & <0.06 & N & Short+EE & L-Merger & \cite{ferro23} \\ 
    191019A & Swift & 64.35$\pm$4.45 & 0.248 & N & <0.01 & N & Long & L-Merger & \cite{levan23b}, \cite{stratta25}, \cite{hjorth05}\\ 
    050724 & Swift & 98.68$\pm$8.56 & 0.257 & N & <0.11 & N & Short+EE & L-Merger & \cite{berger05}, \cite{Malesani+07},\cite{lien16} \\ 
    231117A & Swift & 0.67$\pm$0.07 & 0.257 & N & <0.27 & N & Short & S-Merger & \cite{Schroeder+25}, \cite{2023GCN.35152....1M}\\ 
    120422A & Swift & 60.35$\pm$5.73 & 0.28 & Y & 1.30 & N & Long & Collapsar & \cite{schulze12}\\ 
    150818A & Swift & 143.06$\pm$21.80 & 0.282 & Y & - & N & Long & Collapsar & \cite{lien16}, \cite{Cano+17_review}, \cite{Bi+18} \\ 
    060502B & Swift & 0.14$\pm$0.05 & 0.287 & N & <0.37 & N & Short & S-Merger & \cite{Bloom+07} \\ 
    111225A & Swift & 105.73$\pm$26.18 & 0.297 & A & - & N & Long & L-Weak & \cite{becerra2023understandingnatureopticalemission}\\ 
    050826A & Swift & 29.60$\pm$6.31 & 0.297 & A & - & N & Long & L-Weak & \cite{2005GCN..3888....1M}, \cite{mirabal07} \\ 
    \hline
    \multicolumn{10}{l}{\textbf{\textit{Swift}, $z<0.5$} sample} \\
    \hline
    150727A & Swift & 87.96$\pm$10.99 & 0.313 & A & - & N & Long & L-Weak & \cite{becerra2023understandingnatureopticalemission}\\ 
    130427A & Swift & 244.33$\pm$4.73 & 0.34 & Y & 1.29 & N & Long & Collapsar & \cite{Perley_2014} \\ 
    090417B & Swift & 266.94$\pm$35.39 & 0.345 & A & - & N & Long & L-Weak & \cite{Holland_2010}\\ 
    061021A & Swift & 47.82$\pm$5.63 & 0.346 & A & - & N & Long & L-Weak & \cite{2006GCN..5748....1G}, \cite{Arabsalmani_2017}\\ 
    060428B & Swift & 96.00$\pm$50.59 & 0.348 & A & - & N & Long & L-Weak & \cite{2006GCN..5027....1L}, \cite{lien16}, \cite{Bi+18}\\ 
    130925A & Swift & 160.30$\pm$3.39 & 0.348 & A & - & N & Ultra-Long & L-Weak & \cite{Schady_2015}, \cite{becerra2023understandingnatureopticalemission} \\
    140903A & Swift & 0.296$\pm$0.034 & 0.351 & N & <1.34 & N & Short & S-Weak &  \cite{rastinejad21}, \cite{fong22} \\ 
    130603B & Swift & 0.176$\pm$0.024 & 0.356 & N & <0.01 & Y & Short & S-Merger & \cite{tanvir13}, \cite{fong22}\\ 
    071227A & Swift & 142.48$\pm$48.37 & 0.381 & N & <0.18 & N & Short+EE & L-Merger & \cite{Avanzo_09}, \cite{fong22}\\ 
    120714B & Swift & 157.31$\pm$23.99 & 0.398 & Y & 1.47 & N & Long & Collapsar & \cite{Klose_2019}, \cite{2012CBET.3200....1K} \\ 
    090515 & Swift & 0.036$\pm$0.016 & 0.403 & N & <0.17 & N & Short & S-Merger &  \cite{2010MNRAS.409..531R}, \cite{fong22}\\ 
    061210A & Swift & 85.23$\pm$13.09 & 0.409 & A & - & N & Short+EE & L-Merger & \cite{zhang09}, \cite{Kaneko_2015}\\ 
    100206A & Swift & 0.13$\pm$0.05 & 0.407 & N & <0.31 & N & Short & S-Merger & \cite{Guelbenzu_2012}, \cite{2010GCN.10381....1V}, \cite{Perley_2012}\\ 
    101213A & Swift & 131.12$\pm$41.96 & 0.414 & A & - & N & Long & L-Weak & \cite{101213a_GCN_GMOS_SF}\\ 
    190114C & Swift & 361.46$\pm$11.74 & 0.42 & Y & 7.62 & N & Long & Collapsar & \cite{Melandri_2022} \\ 
    201015A & Swift & 9.78$\pm$3.47 & 0.426 & Y & 0.51 & N & Long & Collapsar & \cite{Belkin_2023} \\ 
    140129B & Swift & 1.36$\pm$0.21 & 0.43 & N & - & N & Short & S-Weak & \cite{fong22}\\ 
    061006A & Swift & 129.79$\pm$30.68 & 0.438 & A & - & N & Short+EE & L-Merger & \cite{fong22} \\ 
    100625A & Swift & 0.332$\pm$0.037 & 0.452 & A & - & N & Short & S-Weak & \cite{fong22} \\ 
    170428A & Swift & 0.20$\pm$0.07 & 0.453 & A & - & N & Short & S-Weak & \cite{fong22}\\ 
    101224A & Swift & 0.244$\pm$0.042 & 0.454 & A & - & N & Short & S-Weak & \cite{Li_2021}, \cite{fong22} \\ 
    070724A & Swift & 0.432$\pm$0.086 & 0.457 & N & <0.01 & N & Short & S-Merger & \cite{2010MNRAS.404..963K}, \cite{fong22}\\ 
    150120A & Swift & 1.196$\pm$0.154 & 0.46 & A & - & N & Short & S-Weak &  \cite{rastinejad21,fong22}\\ 
    210104A & Swift & 32.05$\pm$0.49 & 0.46 & N & <0.48 & N & Long & L-Merger & \cite{Zhang_2022} \\
    150728A & Swift & 0.832$\pm$0.231 & 0.461 & N & - & N & Short & S-Weak & \cite{fong22} \\ 
    070809 & Swift & 1.30$\pm$0.10 & 0.473 & A & 0.61 & Y & Short & S-Merger & \cite{jin20}, \cite{2010ApJ...725.1202L}, \cite{fong22} \\ 
    130831A & Swift & 30.19$\pm$2.33 & 0.479 & Y & 1.08 & N & Long & Collapsar &
    \cite{Klose_2019} 
    \\ 
    200219A & Swift & 288.0$\pm$50.60 & 0.48 & N & - & N & Short+EE & L-Weak & \cite{fong22} \\
    051117B & Swift & 9.02$\pm$1.32 & 0.481 & A & - & N & Long & L-Weak & \cite{Chandra_2012}, \cite{Schulze_2015} \\ 
    160624A & Swift & 0.192 $\pm$ 0.143 & 0.483 & A & - & N & Short & S-Weak & \cite{2016GCN.19565....1C}, \cite{fong22}\\ 
    091127A & Swift & $6.96 \pm 0.15$ & 0.49 & Y & 2.76 & N & Long & Collapsar & \cite{Cobb_2010}, \cite{Klose_2019} \\ 
    \hline
    \multicolumn{10}{l}{\textbf{All telescopes, $z<0.5$} sample} \\
    \hline
    980425 & BeppoSAX & 34.88$\pm$3.781 & 0.0085 & Y & 1 & N & Long & Collapsar &  \cite{galama98,savaglio09,Clocchiatti+11} \\
    170817A & Fermi & $2.0 \pm 0.5$ & 0.0098 & N & <0.01 & Y & Short & S-Merger & \cite{goldstein17,Levan+17,villar18} \\ 
    230307A & Fermi & 34.56$\pm$0.57 & 0.065 & N & <0.01 & Y & Long & L-Merger & \cite{levan24a} \cite{yang23} \\ 
    031203A & INTEGRAL & 20$\pm$0 & 0.1 & Y & 2.04 & N & Long & Collapsar & \cite{2004ApJ...608L..93c}, \cite{Klose_2019} \\ 
    030329 & HETE-2 & 22.9$\pm$0 & 0.169 & Y & 1.08 & N & Long & Collapsar & \cite{Modjaz_2016}, \cite{Klose_2019} \\ 
    130702A & Fermi & 58.88$\pm$6.19 & 0.145 & Y & 1.22 & N & Long & Collapsar & \cite{Volnova_2017}\\ 
    050709 & HETE-2 & 0.22$\pm$0.05 & 0.16 & N & <0.01 & Y & Short & S-Merger &  \cite{hjorth05a}, \cite{Covino_2006}, \cite{Fox_2005} \\ 
    040701 & HETE-2 & 60$\pm$0 & 0.2 & N & <0.01 & N & Long & L-Merger & \cite{2005ApJ...627..877S} \\ 
    020903A & HETE-2 & 3.3$\pm$0 & 0.3 & Y & 0.52 & N & Long & Collapsar & \cite{2006ApJ...643..284B}, \cite{2005ApJ...627..877S}, \cite{2002GCN..1530....1R}\\ 
    220219B & Fermi & <79.0 & 0.293 & Y & 2.13 & N & Long & Collapsar & \cite{2022GCN.31619....1H}\\
    150518A & MAXI & 250 & 0.256 & Y & - & A & Long & Collapsar & \cite{2015GCN.17903....1P}, \cite{Cano+17}\\ 
    171010A & Fermi & 107.27$\pm$0.81 & 0.33 & Y & 1.55 & N & Long & Collapsar & \cite{Melandri+19} \\ 
    230812B & Fermi & 2.95$\pm$1.02 & 0.36 & Y & 0.97 & N & Long & Collapsar & \cite{Srinivasaragavan_2024} \\
    011121A & BeppoSAX & 28 & 0.36 & Y & 0.55 & N & Long & Collapsar & \cite{bloom02}\\ 
    160623A & Fermi & 107.78$\pm$8.69 & 0.367 & A & - & A & Long & L-Weak & \cite{2016GCN.19708....1M} \\
    140606B & Fermi & 22.78$\pm$2.06 & 0.384 & Y & 0.87 & N & Long & Collapsar & \cite{Cano_2015} \\ 
    211023A & Fermi & 79.11$\pm$0.57 & 0.39 & Y & 1.83 & N & Long & Collapsar &  \cite{2021GCN.31098....1B}, \cite{2021GCN.31053....1P}\\ 
    990712B & BeppoSAX & 19.0$\pm$4.5 & 0.433 & Y & 0.5 & N & Long & Collapsar & \cite{2000ApJ...534L.147H}, \cite{Bjornsson+01}, \cite{Frontera_2009} \\ 
    010921 & HETE-2 & 34.2 & 0.451 & N & <1.45 & A & Long & L-Weak & \cite{Price_2003} \\ 
    111211A & AGILE & 15 & 0.478 & Y & 1.70 & N & Long & Collapsar & \cite{2012GCN.12802....1D}, \cite{kruehler15}\\  
    \hline
\end{xltabular}

%% file: F_98s_HERE.tex
\begin{table}
\centering
\footnotesize 
\caption{Observations used infer $f_{1998bw}$. Data references are compiled in Table \ref{sample}.\\
$^{1}\Delta T$ denotes the observed-frame time between the GRB detection and the observation. \\  
$^{2}$All observations are corrected for Milky Way extinction \citep{SF11} before calculating $f_{\rm 1998bw}$. We mark a $\star$ where the observation was corrected for Milky Way extinction in the source from which it is drawn and therefore not applied here. We record a value where we applied the correction. \\  
$^{3}$Values for additional line-of-sight extinction where applied. These are described in more detail in Appendix~\ref{individualGRBs}.}\label{sample_photometry}
\begin{tabular}{llccccccccc}
\hline
GRB & $z$ & SN? & Category & $\Delta T^{1}$ & Band & Brightness or Upper Limit & $A^{2}_{\text{MW}}$ & E(B-V)$^{2}_{\rm MW}$ & E(B-V)$^{3}_{\rm LOS}$ & $f_{\rm 1998bw}$  \\
 &  &  &  & (days) &  & (AB mag) & (mag) & (mag) & (mag) &   \\
\hline
\multicolumn{11}{l}{\textbf{\textit{Swift}, $z<0.3$} sample} \\
\hline
111005A & 0.013 & N & L-Merger & 0.65 & I & >19.5 & 0.255 & 0.082 & - & <0.02\\
060218 & 0.0331 & Y & Collapsar & 15.96 & R & 17.1 & 0.406 & 0.131 & - & 0.85 \\
171205A & 0.0368 & Y & Collapsar & 16.01 & i' & 18.2 & 0.138 & 0.045 & - & 0.63 \\
100316D & 0.0591 & Y & Collapsar & 15.49 & r & 20.1 & 0.372 & 0.12 & 0.14 - 0.90 & 0.30 - 2.10 \\
190829A & 0.0785 & Y & Collapsar & 14.30 & i & 20.8 & $\star$ & $\star$ & 0.64 - 1.04 & 0.74- 1.64 \\
211211A & 0.076 & N & L-Merger & 46.95 & r & >24.5 & $\star$ & $\star$ & - & <0.01 \\
051109B & 0.08 & N & L-Weak & 24.74 & R & >20.6 & 0.45 & 0.145 & - & <0.26 \\
060505 & 0.089 & N & L-Merger & 18. & R & >25.2 & 0.056 & 0.018 & - & <0.01 \\
180728A & 0.117 & Y & Collapsar & 16.32 & z' & 20.4 & 0.756 & 0.244 & - & 0.93 \\
080905A & 0.122 & N & S-Merger & 1.5 & R & >25.4 & $\star$& $\star$& - & <0.03 \\
060614 & 0.13 & N & L-Merger & 7.84 & I & >25.6 & $\star$& $\star$& - & <0.02 \\
150101B & 0.134 & N & S-Merger & 10.71 & r & >24.2 & $\star$& $\star$& - & <0.02 \\
161219B & 0.147 & Y & Collapsar & 15.37 & r & 19.7 & $\star$& $\star$& - & 1.17 \\
221009A & 0.151 & Y & Collapsar & 18.31 & r & 23.0 & 4.22 & 1.36 & - & 1.37 \\
130822A & 0.154 & N & S-Weak & 6.69 & J & >22.9 & $\star$& $\star$& - & <0.16 \\
160821B & 0.162 & N & S-Merger & 10 & g & >25.7 & $\star$& $\star$& - & <0.011 \\
050509B & 0.225 & N & S-Merger & 2.67 & R & >26.2 & $\star$& $\star$& - & <0.04\\
211227A & 0.228 & N & L-Merger & 1.24 & r & >25.8 & 0.055 & 0.018 & - & <0.06 \\
191019A & 0.248 & N & L-Merger & 37.41 & r & >26.8 & $\star$& $\star$& - & <0.01 \\
050724 & 0.257 & N & L-Merger & 3.46 & I & 25.6 & $\star$& $\star$& - & <0.11 \\
231117A & 0.257 & N & S-Weak & 5.91 & r & >23.6 & 0.196 & 0.063 & - & <0.27 \\
120422A & 0.28 & Y & Collapsar & 17.59 & r' & 21.2 & $\star$ & $\star$ & - & 1.30 \\
060502B & 0.287 & N & S-Merger & 0.7 & R & >24.5 & $\star$ & $\star$ & - & <0.37 \\
\hline
\multicolumn{11}{l}{\textbf{\textit{Swift}, $z<0.5$} sample} \\
\hline
130427A & 0.34 & Y & Collapsar & 16.73 & r & 21.8 & 0.055 & 0.018 & - & 1.29 \\
140903A & 0.351 & N & S-Weak & 7.53 & r & >22.6 & 0.089 & 0.029 & - & <1.34 \\
130603B & 0.356 & N & S-Merger & 9.37 & F606W & >28.2 & 0.063 &  & - & <0.01 \\
071227A & 0.381 & N & L-Merger & 21.25 & R & >24.1 & $\star$ & $\star$ & - & <0.18 \\
120714B & 0.398 & Y & Collapsar & 15.51 & r' & 22.3 & 0.248 & 0.08 & - &1.47 \\
090515 & 0.403 & N & S-Merger & 1.04 & r & >26.5 & 0.058 & 0.019 & - &<0.17 \\
100206A & 0.407 & N & S-Weak & 0.65 & i & >25.8 & $\star$ & $\star$ & - &<0.31 \\
190114C & 0.42 & Y & Collapsar & 19 & i & 20.9 & 0.035 & 0.013 & 0.30 & 7.62 \\
201015A & 0.426 & Y & Collapsar & 19.74 & R & 23.3 & 0.93 & 0.018 & - & 0.51 \\
070724A & 0.457 & N & S-Merger & 18.14 & R & >27.4 & 0.0403 & 0.013 & - & <0.01 \\
210104A & 0.46 & N & L-Weak & 11.46 & R & >23.9 & $\star$ & $\star$ & - & <0.48 \\
070809 & 0.473 & A & S-Weak & 1.47 & R & 25.5 & 0.248 & 0.08 & - & 0.61 \\
130831A & 0.479 & Y & Collapsar & 17.56 & z & 22.7 & 0.124 & 0.04 & - & 1.08 \\
091127A & 0.49 & Y & Collapsar & 18.06 & I & 21.6 & $\star$ & $\star$ & - & 2.76 \\
\hline
\multicolumn{11}{l}{\textbf{All telescopes, $z<0.5$} sample} \\
\hline
170817A & 0.0098 & N & S-Merger & 18.43 & $K_s$ & >22.7 & 0.191 & 0.062 & - & <0.01 \\
980425 & 0.0085 & Y & Collapsar & 15.35 & B & 14.1 & 0.16 & 0.052 & - & 1.0 \\
230307A & 0.065 & N & L-Merger & 28.89 & F070W   & >29.0 & 0.239 & 0.077 & - & <0.01 \\
031203A & 0.1 & Y & Collapsar & 16 & J & 19.0 & 2.42 & 0.78 & - & 2.04\\
030329 & 0.17 & Y & Collapsar & 19.70 & R & 20.2 & 0.069 & 0.022 & - & 1.08 \\
130702A & 0.145 & Y & Collapsar & 16.67 & R & 19.8 & 0.118 & 0.038 & - & 1.22 \\
050709 & 0.16 & N & S-Merger & 9.8 & F814W & >25.8 & $\star$ & $\star$ & - & <0.01 \\
040701 & 0.215 & N & L-Merger & 39.1 & F625W & >27.8 & 0.403 & 0.13 & - & <0.01 \\
020903A & 0.251 & Y & Collapsar & 26.50 & R & 21.9 & 0.102 & 0.033 & - & 0.52 \\
220219B & 0.293 & Y & Collapsar & 8.45 & r & 21.2 & $\star$ & $\star$ & - & 2.13 \\
171010A & 0.33 & Y & Collapsar & 17.48 & r & 20.2 & 0.408 & 0.13 & - & 1.55 \\
230812B & 0.36 & Y & Collapsar & 18.43 & r & 22.2 & 0.062 & 0.02 & - & 0.97 \\
011121A & 0.36 & Y & Collapsar & 23.03 & V & 24.4 & 1.34 & 0.432 & - & 0.55 \\
140606B & 0.384 & Y & Collapsar & 17.01 & i & 22.3 & $\star$ & $\star$ & - & 0.87 \\
211023A & 0.39 & Y & Collapsar & 16.9 & R & 21.7 & $\star$ & $\star$ & - & 1.83 \\
990712B & 0.433 & Y & Collapsar & 47.7 & V & 25.6 & $\star$ & $\star$ & - & 0.5 \\
010921 & 0.451 & N & L-Weak & 28.05 & R & >24.2 & 0.398 & 0.128 & 0.37 & <1.45 \\
111211A & 0.478 & Y & Collapsar & 15.3 & r & 22.6 & 0.131 & 0.05 & - & 1.70 \\
\hline \hline
\end{tabular}
\end{table}

%% file: WRAPPED_host_prop_table.tex
\setlength{\tabcolsep}{3pt} 
\captionof{table}{Properties, where available, of the host galaxies and afterglows of GRBs in our sample and shown in Figure~\ref{gold_sample_box}. \\ 
$^{\dagger}$Host type abbreviations are: ``SF'' - star-forming, ``Q'' - quiescent, ``T'' - transitioning, ``U'' - unknown. \\ 
Values of $n_{H}$ and $F_{\rm X, 11hr}$, where available, are taken from UKSSDC \citep{evans07,evans09}.
}
\label{tab:host_props}
\begin{xltabular}{\textwidth}{l c c c c c c X}
    \hline
    GRB & Host Type$^{\dagger}$ & Host SFR & Host Stellar Mass & Offset & $n_{H}$ & $F_{\rm X, 11hr}$ & Host Galaxy or Offset Refs. \\
     &  & log($M_{\odot}$) & log($M_{\odot}$) & '' & cm$^{-3}$ & $10^{-11}$ erg s$^{-1}$cm$^{-2}$ & \\
    \hline \endhead
    \hline
    \multicolumn{8}{r}{\textit{Table continued on next page}} \\
    \endfoot

    \hline
    \endlastfoot
    \multicolumn{8}{l}{\textbf{\textit{Swift}, $z<0.3$} sample} \\
    \hline
     111005A & SF & 0.38 & $9.7 \pm -99.0$ & $0.50 \pm 0.50$ &  &  & \cite{michalowski18} \\
    100316D & SF & $1.7 \pm 0.1$ & $8.9 \pm -99.0$ &  & $5.60 \pm 17.90$ & $0.08$ & \cite{starling11},This work \\
    060218 & SF & $0.1 \pm 0.0$ & $7.2 \pm 0.3$ & $0.17 \pm 0.04$ & $4.40 \pm 0.60$ & $0.04$ & \cite{blanchard16} \\
    171205A & SF & $3.0 \pm 1.0$ & $10.1 \pm 0.1$ & $7.15 \pm 0.01$ & $0.15 \pm 0.54$ & $0.08$ & \cite{izzo19,thoene24} \\
    190829A & T & $6.9 \pm 0.5$ & $12.8 \pm 0.0$ & $9.55 \pm 0.01$ & $13.80 \pm 1.00$ & $2.31$ & \cite{gupta22,bhi24} \\
    211211A & SF & $0.1 \pm 0.0$ & $8.8 \pm 0.1$ & $5.44 \pm 0.02$ & $0.05 \pm 0.27$ & $0.48$ & \cite{fong22,nugent22} \\
    051109B & SF &  &  & $14.99 \pm 1.60$ & $0.90 \pm 0.90$ & $0.02$ & \cite{perley06} \\
    060505 & SF & 0.43 & $9.4 \pm 0.0$ & $4.31 \pm 0.04$ & $0.95 \pm 2.52$ & $0.08$ & \cite{blanchard16,thoene08} \\
    080517 & SF & $15.5 \pm 0.4$ & $9.6 \pm 0.1$ & $3.60 \pm 1.80$ & $2.90 \pm 2.40$ & $0.01$ & \cite{stanway15} \\
    150101B & Q & $0.2 \pm 0.0$ & $11.1 \pm 0.0$ & $3.07 \pm 0.03$ &  &  & \cite{fong22,nugent22} \\
    180728A & SF &  &  & $0.1 \pm 0.1$ & $0.61 \pm 0.28$ & $3.45$ & \cite{Rossi+26} \\
    080905A & SF & 0.3 & $10.3 \pm 0.5$ & $8.20 \pm 0.30$ & $2.30 \pm 2.60$ & $0$ & \cite{rowlinson10,rossi20} \\
    060614 & Q & $0.0 \pm 0.0$ & $7.8 \pm 0.0$ & $0.36 \pm 0.01$ & $0.14 \pm 0.24$ & $0.55$ & \cite{blanchard16} \\
    161219B & SF & $0.2 \pm 0.2$ & $8.9 \pm 0.5$ & $1.50 \pm 0.20$ & $1.94 \pm 0.25$ & $1.46$ & \cite{Cano+17} \\
    221009A & SF & 0.17 & $9.6 \pm 0.1$ & $0.30 \pm 0.01$ & $14.00 \pm 4.00$ &  & \cite{levan23a,blanchard23} \\
    130822A & SF & $2.2 \pm 0.4$ & $10.2 \pm 0.1$ & $22.38 \pm 1.82$ & $0.80 \pm 5.90$ &  & \cite{fong22,nugent22} \\
    160821B & SF & $0.2 \pm 0.0$ & 9.24 & $5.61 \pm 0.01$ & $0 \pm 0.45$ & $0$ & \cite{fong22,nugent22} \\
    050219A & Q & 0.06 & 9.98 & $2.00 \pm 2.40$ & $1.40 \pm 0.70$ & $0.07$ & \cite{rossi14} \\
    050509B & Q & $0.2 \pm 0.0$ & 11.46 & $15.10 \pm 3.40$ & $0 \pm 3.14$ & $0$ & \cite{fong22,nugent22} \\
    120305A & T & 0.03 & $9.2 \pm 0.1$ & $4.97 \pm 0.04$ & $5.10 \pm 4.00$ & $0$ & \cite{fong22,nugent22} \\
    211227A & T & $0.7 \pm 0.0$ & $10.5 \pm 0.0$ & $3.76 \pm 2.40$ & $3.50 \pm 2.10$ & $0$ & \cite{ferro23} \\
    191019A & Q & $0.1 \pm 0.0$ & $10.5 \pm 0.1$ & $0.02 \pm 0.03$ & $1.20 \pm 1.60$ & $0.01$ & \cite{levan23b} \\
    050724 & Q & $0.1 \pm 0.0$ & $11.1 \pm 0.0$ & $0.68 \pm 0.02$ & $0 \pm 1.27$ & $0.01$ & \cite{Bi+18,fong22,nugent22} \\
    231117A & SF & $0.4 \pm 0.0$ & $9.2 \pm 0.1$ & $0.43 \pm 0.02$ & $0 \pm 0.07$ & $0.07$ & \cite{Schroeder+25} \\
    120422A & SF & $0.4 \pm 0.1$ & $8.9 \pm 0.0$ & $1.70 \pm 0.10$ & $0 \pm 1.05$ & $0.01$ & \cite{schulze12} \\
    150818A & SF & 0 &  &  & $0.60 \pm 0.80$ & $0.05$ & \cite{Taggart+21} \\
    060502B & Q & 0.8 & 11.84 & $17.05 \pm 4.36$ & $2.00 \pm 8.20$ & $0$ & \cite{Bloom+07} \\
    111225A & SF &  &  &  & $0 \pm 1.49$ & $0.01$ & \cite{Taggart+21} \\
    050826A & SF & 9.13 & $9.8 \pm 0.1$ & $0.40 \pm 0.05$ & $10 \pm 4.00$ & $0.02$ & \cite{mirabal07} \\
    \hline
    \multicolumn{8}{l}{\textbf{\textit{Swift}, $z<0.5$} sample} \\
    \hline
    150727A & SF &  &  &  & $0 \pm 1.02$ & $0.03$ & \cite{150727a_gcn} \\
    130427A & SF & $23.1 \pm 0.0$ & $9.6 \pm 0.0$ & $0.83 \pm 0.01$ & $1.10 \pm 0.40$ & $9.83$ & \cite{Perley_2014,Levan_130427a} \\
    090417B & SF & 1 & $10.1 \pm 0.1$ &  & $33.00 \pm 3.00$ & $0.47$ & \cite{Holland_2010,savaglio09}  \\
    061021A & SF & 0.05 & $8.5 \pm 0.5$ &  & $0.58 \pm 0.27$ & $0.32$ & \cite{Arabsalmani18,kruehler15,kruhler17}  \\
    060428B & Q &  &  &  & $0 \pm 0.18$ & $0.03$ & \cite{gcn_060428b} \\
    130925A & SF & $18.2 \pm 0.1$ & $9.5 \pm 0.0$ & $0.16 \pm 0.01$ & $30.80 \pm 3.00$ & $1.93$ & \cite{Schady15} \\
    140903A & SF & $2.3 \pm 0.4$ & $10.8 \pm 0.0$ & $0.18 \pm 0.02$ & $1.80 \pm 0.90$ & $0.12$ & \cite{fong22,nugent22} \\
    130603B & SF & $0.4 \pm 0.1$ & $9.8 \pm 0.0$ & $1.07 \pm 0.04$ & $4.33 \pm 1.03$ & $0.04$ & \cite{fong22,nugent22} \\
    071227A & SF & $5.8 \pm 4.3$ & $10.5 \pm 0.1$ & $2.80 \pm 0.05$ & $1.10 \pm 2.10$ & $0$ & \cite{Avanzo_09} \\
    120714B & SF & $0.5 \pm 0.2$ & $8.7 \pm 0.2$ & $0.21 \pm 0.20$ & - & $0$ & \cite{Klose_2019} \\
    090515 & Q & 0 & $11.2 \pm 0.1$ & $13.98 \pm 0.03$ & $0.27 \pm 0.28$ & $0$ & \cite{fong22,nugent22} \\
    061210A & SF & $0.2 \pm 0.2$ & $9.5 \pm 0.0$ & $2.82 \pm 2.61$ & $9.90 \pm 16.80$ & $0.16$ & \cite{fong22,nugent22} \\
    100206A & SF & $4.6 \pm 0.1$ & $10.7 \pm 0.0$ & $4.59 \pm 2.37$ & $4.10 \pm 11.50$ & $0$ & \cite{fong22,nugent22} \\
    101213A & SF &  &  &  & $9.80 \pm 1.40$ & $0.18$ & \cite{101213a_GCN_GMOS_SF} \\
    190114C & SF & $9.4 \pm 9.6$ & $9.3 \pm 0.3$ & $0.03 \pm 0.01$ & $79.00 \pm 6.00$ & $1.12$ & \cite{deUgartePostigo_190114C,Melandri_2022} \\
    201015A & U &  &  & $1.50 \pm 0.10$ & $6.00 \pm 7.00$ & $0.01$ & \cite{Belkin_2023} \\
    140129B & SF & $0.1 \pm 0.0$ & $9.3 \pm 0.1$ & $0.31 \pm 0.31$ & $6.00 \pm 6.00$ & $0.02$ & \cite{fong22,nugent22} \\
    061006A & T & $0.1 \pm 0.1$ & $9.4 \pm 0.1$ & $0.24 \pm 0.05$ & $0 \pm 1.05$ & $0.02$ & \cite{fong22,nugent22} \\
    100625A & Q & 0 & $9.7 \pm 0.1$ & $0.45 \pm 1.16$ & $0 \pm 1.04$ & $0$ & \cite{fong22,nugent22} \\
    170428A & SF & $0.4 \pm 0.0$ & $9.7 \pm 0.0$ & $1.32 \pm 0.58$ & $3.60 \pm 4.50$ & $0$ & \cite{fong22,nugent22} \\
    101224A & SF & $0.6 \pm 0.1$ & $9.2 \pm 0.1$ & $2.08 \pm 2.31$ & $11.00 \pm 67.00$ & $0$ & \cite{fong22,nugent22} \\
    070724A & SF & $6.5 \pm 0.1$ & 9.81 & $0.94 \pm 0.03$ & $1.50 \pm 2.60$ & $0$ & \cite{2010MNRAS.404..963K,fong22,nugent22} \\
    150120A & SF & $2.3 \pm 0.8$ & $10.0 \pm 0.1$ &  & $7.00 \pm 5.00$ &  & \cite{fong22,nugent22} \\
    210104A & U &  &  &  & $7.50 \pm 3.10$ & $0.48$ &  \\
    150728A & SF & $8.1 \pm 0.6$ & $9.3 \pm 0.0$ & $1.28 \pm 3.44$ & $150 \pm 220$ &  & \cite{fong22,nugent22} \\
    070809 & T & $0.8 \pm 0.7$ & $10.8 \pm 0.0$ & $5.70 \pm 0.46$ & $0 \pm 0.31$ & $0.05$ & \cite{fong22,nugent22} \\
    130831A & SF & $0.6 \pm 0.3$ & $8.4 \pm 0.4$ & $0.92 \pm 0.20$ & $0.06 \pm 0.59$ & $0.23$ & \cite{Klose_2019} \\
    200219A & SF & $9.9 \pm 1.6$ & $10.7 \pm 0.0$ & $1.38 \pm 0.88$ & $0 \pm 0.79$ & $0$ & \cite{fong22,nugent22} \\
    051117B & SF & $4.4 \pm 0.1$ & $10.2 \pm 0.2$ &  & $2.50 \pm 3.00$ & $0$ & \cite{kruehler15,schady07,kruhler17}  \\
    160624A & SF & $1.3 \pm 0.8$ & $9.7 \pm 0.1$ & $1.59 \pm 1.03$ & $2.80 \pm 2.80$ & $0$ & \cite{fong22,nugent22} \\
    091127A & SF & $0.5 \pm 0.2$ & $8.8 \pm 0.0$ & $0.22 \pm 0.05$ & $1.10 \pm 0.60$ & $2.05$ & \cite{Klose_2019,kruhler17} \\
    \hline
    \multicolumn{8}{l}{\textbf{All telescopes, $z<0.5$} sample} \\
    \hline
    170817A & Q &  &  &  &  &  & \cite{Blanchard+17,Levan+17} \\
    980425 & SF & $0.3 \pm 0.1$ & $8.7 \pm 0.3$ &  &  &  & \cite{savaglio09} \\
    230307A & SF & $0.4 \pm 0.3$ & $9.4 \pm 0.1$ &  &  &  & \cite{levan24a} \\
    031203A & SF & 12.68 & $8.8 \pm 0.4$ &  &  &  & \cite{Klose_2019} \\
    030329 & SF & 0.11 & $7.7 \pm 0.1$ &  &  &  & \cite{Klose_2019} \\
    130702A & SF & 0.05 & 8.11 &  &  &  & \cite{Klose_2019,Volnova_2017}. \\
    050709 & SF &  &  &  &  &  & \cite{fong22,nugent22} \\
    040701 & SF &  &  &  &  &  & \cite{Soderberg_HSTSNe} \\
    020903A & SF & 2.65 & $8.9 \pm 0.1$ &  &  &  & \cite{Soderberg_HSTSNe} \\
    150518A & SF &  &  &  &  &  & \cite{grb150518a_vlt} \\
    171010A & SF & $1.1 \pm 0.2$ &  & $1.40 \pm 0.10$ &  &  & \cite{Melandri+19} \\
    230812B & SF &  &  &  &  &  & \cite{Srinivasaragavan_2024} \\
    011121A & SF & 2.24 & $9.8 \pm 0.2$ &  &  &  & \cite{Bloom+02_011121} \\
    160623A & SF &  &  &  &  &  & \cite{grb150518a_vlt} \\
    140606B & SF & $0.1 \pm 0.0$ &  &  &  &  & \cite{Klose_2019,Cano_2015} \\
    211023A & SF &  &  &  &  &  & \cite{fong22,nugent22} \\
    990712B & SF & 2.39 & $9.3 \pm 0.0$ &  &  &  & \cite{Bjornsson+01,Frontera_2009} \\
    010921 & SF & 2.5 & $9.7 \pm 0.1$ &  &  &  & \cite{Price_2003} \\
    111211A & SF & $13.6 \pm -0.0$ & $9.1 \pm 0.0$ &  &  &  &  \cite{kruehler15,kruhler17} \\
    \hline
\end{xltabular}